\documentclass[%
aps,
twocolumn,prd,
showpacs,preprintnumbers,nofootinbib,
amsmath,amssymb,
aps,
superscriptaddress]{revtex4-1}

\usepackage{float}
\usepackage{graphicx}
\usepackage{dcolumn}
\usepackage{bm}
\usepackage[colorlinks=true]{hyperref}
\usepackage{color}
\usepackage{array}
\usepackage{multirow}
\usepackage[normalem]{ulem}

\definecolor{RED}{rgb}{1,0,0}

\def\rchisq{\chi_r^2}
\def\pycbc{PyCBC}
\def\phenomd{\texttt{IMRPhenomD}}

\begin{document}

\title{Sensitivity of gravitational wave searches to the full signal of intermediate mass black hole binaries during the LIGO O1 Science Run}

\affiliation{Center for Relativistic Astrophysics and School of Physics, Georgia Institute of Technology, Atlanta, GA 30332}
\affiliation{Albert Einstein Institute (Max Planck Institut f\"ur Gravitationsphysik), Callinstrasse 38, Hannover, Germany}
\affiliation{NASA Postdoctoral Program Fellow, Goddard Space Flight Center, Greenbelt, MD 20771, USA}

\author{Juan Calder\'on~Bustillo$^\text{1}$}\noaffiliation
\author{Francesco Salemi$^\text{2}$}\noaffiliation
\author{Tito \surname{Dal Canton}$^\text{3}$}\noaffiliation
\author{Karan P.~Jani$^\text{1}$}\noaffiliation


\preprint{LIGO-P1700204}
\pacs{04.80.Nn, 04.25.dg, 04.25.D-, 04.30.-w}

\begin{abstract}
The sensitivity of gravitational wave searches for binary black holes is estimated via the injection and posterior recovery of simulated gravitational wave signals in the detector data streams. When a search reports no detections, the estimated sensitivity is then used to place upper limits on the coalescence rate of the target source. In order to obtain correct sensitivity and rate estimates, the injected waveforms must be faithful representations of the real signals. Up to date, however, injected waveforms have neglected radiation modes of order higher  than the quadrupole, potentially biasing sensitivity and coalescence rate estimates. In particular, higher-order modes are known to have a large impact in the gravitational waves emitted by intermediate mass black holes binaries. In this work we evaluate the impact of this approximation in the context of two search algorithms run by the LIGO Scientific Collaboration in their search for intermediate mass black hole binaries in the O1 LIGO Science Run data: a matched-filter based pipeline and a coherent un-modeled one. To this end we estimate the sensitivity of both searches to simulated signals including and omitting higher-order modes. We find that omission of higher-order modes leads to biases in the sensitivity estimates which depend on the masses of the binary, the search algorithm and the required level of significance for detection. In addition, we compare the sensitivity of the two search algorithms across the studied parameter space. We conclude that the most recent LIGO-Virgo upper limits on the rate of coalescence of intermediate mass black hole binaries are conservative for the case of highly asymmetric binaries. However, the tightest upper limits, placed for nearly-equal-mass sources, remain unchanged due to the small contribution of higher modes to the corresponding sources.
\end{abstract}

\maketitle


\section{Introduction}
The scientific community has recently witnessed the birth of the field of gravitational wave (GW) astronomy. In the first half of 2016, the LIGO and Virgo Scientific Collaborations announced the detection of two GW signals produced by the mergers of binary black holes (BBH), GW150914 and GW151226 \cite{Abbott:2016blz,Abbott:2016nmj,TheLIGOScientific:2016pea}. A third BBH merger, GW170104, was detected in the first months of the O2 Science run of Advanced LIGO \cite{Abbott:2017vtc}. This was recently followed by the detection of GW170814 two weeks before the end of the O2 Science run \cite{Abbott:2017oio}. All of these sources have the common properties of being BBHs with a total mass below $100M_\odot$, relatively low mass ratios and no clear evidence of a precessing orbital plane. BBHs heavier than $100M_\odot$ are commonly known as intermediate mass black hole binaries (IMBHB) \cite{Ferrarese:2000se,Gebhardt:2000fk,Graham:2010nb,Graham:2012pw,Gultekin:2009bm,Marconi:2003hj,McConnell:2012hz}. During the O1 LIGO Science run, a dedicated search targeting such sources reported no detections and placed upper limits of their rate of coalescence \cite{Abbott:2017iws}. Notably, these limits improved by a factor of 100 with respect to previous studies \cite{Aasi:2014iwa}. In particular, an upper limit of $0.94$ Gpc$^{-3}yr^{-1}$ was achieved for BBH mergers with a total mass of $200M_{\odot}$.

Upper limits on coalescence rates are computed in the following way \cite{Abbott:2017iws}. First, sets of simulated GW signals are injected in the detectors data streams. Next, searches for GWs are run on such injection sets to estimate the sensitivity of the searches to the corresponding sources. In order to obtain accurate results, it is crucial that GW signals are simulated with accurate models of the real signal emitted by the source that is being considered. In fact, injecting inaccurate signals would bias the corresponding sensitivity and upper limit estimates.

The GW signal $h$ emitted by a BBH can be decomposed as a superposition of several GW modes $h_{\ell,m}$. Past studies of upper limits have approximated the real GW signal using the dominant quadrupolar $(\ell,m)=(2,\pm 2)$ mode only, omitting the impact of higher-order modes \cite{Abbott:2017iws}. However, the latter are known to have a large contributions to the GW signal when the total mass $M=m_1+m_2$ of the system is above $\sim 100 M_{\odot}$, which is the case of IMBHB. Here, $m_1$ and $m_2$ are the component masses of the binary and we will always consider $m_1 \geq m_2$. Although the impact of higher-order modes is small for equal-mass systems, it grows with the mass ratio $q=m_1/m_2 \geq 1$ \cite{Bustillo:2015qty,Bustillo:2016gid, Capano:2013raa,Varma:2014jxa,Pekowsky:2012sr}. Therefore, ignoring higher-order modes for unequal-mass systems will generally lead to biased sensitivity and rates estimates.

The main goal of this work is to quantify the impact of the omission of higher-order modes on sensitivity estimations of the GW search for IMBHB during the O1 LIGO Science Run presented in \cite{Abbott:2017iws}. We focus on two of the analysis methods employed in that study: the matched-filter based pipeline \pycbc{} \cite{Canton:2014ena,Usman:2015kfa,pycbc-software} and the un-modeled Coherent Wave Burst pipeline $\verb+cWB+$ \cite{Klimenko:2008fu}. To this end, we inject simulated IMBHB signals including and omitting higher order modes in the O1 Science Run data, and estimate the sensitivity of the two searches to both kinds of signals. By studying how the sensitivity changes when higher-order modes are included, we asses the accuracy of the rate limits presented in \cite{Abbott:2017iws}. We stress that we run both pipelines in the same configuration used for the IMBHB search in the LIGO O1 Science Run.

As we go to higher and higher masses, deep into the IMBHB region, compact binary waveforms observed by ground-based interferometers eventually lose their characteristic ``chirp'' morphology and only consist of a few oscillations, becoming more difficult to distinguish from instrumental transients. It is therefore interesting to ask whether a matched-filter analysis tuned to compact-binary-merger signals continues to be more sensitive than an analysis which is nearly agnostic about the signal's time-frequency morphology. Using the same simulations described above, we also investigate this question.

The rest of this paper is organized as follows. Section II reviews the concept of higher-order modes and describes the source configurations for which such modes are important. Section III describes the two search algorithms we consider in this study, the matched-filter based \pycbc{} and the un-modeled search known as Coherent Wave Burst ($\verb+cWB+$). Section IV illustrates how upper limits on coalescence rate are estimated via signal injections and describes the waveform models we use as a representation of the true GW signal. In section V we report and compare the sensitivity of the two studied pipelines to our simulated signals. Section VI discusses the impact of these results in the latest IMBHB coalescence rates published by LIGO and Virgo and in section VIII we report our conclusions. In addition, we attach an appendix addressing the accuracy of the waveform model we use as a representation of the true GW signal.
 
\section{The BBH signals: Higher modes}
The GW emission from a non-eccentric BBH depends on 15 parameters. The masses $m_i$ and dimensionless spins $\vec\chi_i=\vec{s}_i/m_i$ of its two component bodies are known as the intrinsic parameters and we will collectively denote these by $\Xi$. Consider a frame of reference in standard spherical coordinates $(r,\iota,\varphi)$ with origin in the center of mass of the BBH. The polar angle $\iota$ is defined such that $\iota=0$ coincides with the total angular momentum of the binary. The remaining 4 parameters are the time of coalescence $t_c$, the sky-location $(\bar\theta,\bar\varphi)$, and the polarisation $\psi_p$ of the signal. The latter three impact the observed signal through the so-called antenna pattern $(F_+,F_\times)$ of the detector \cite{Jaranowski:1998qm,Varma:2014jxa}.

\begin{figure}
\centering
\includegraphics[width=1.0\columnwidth]{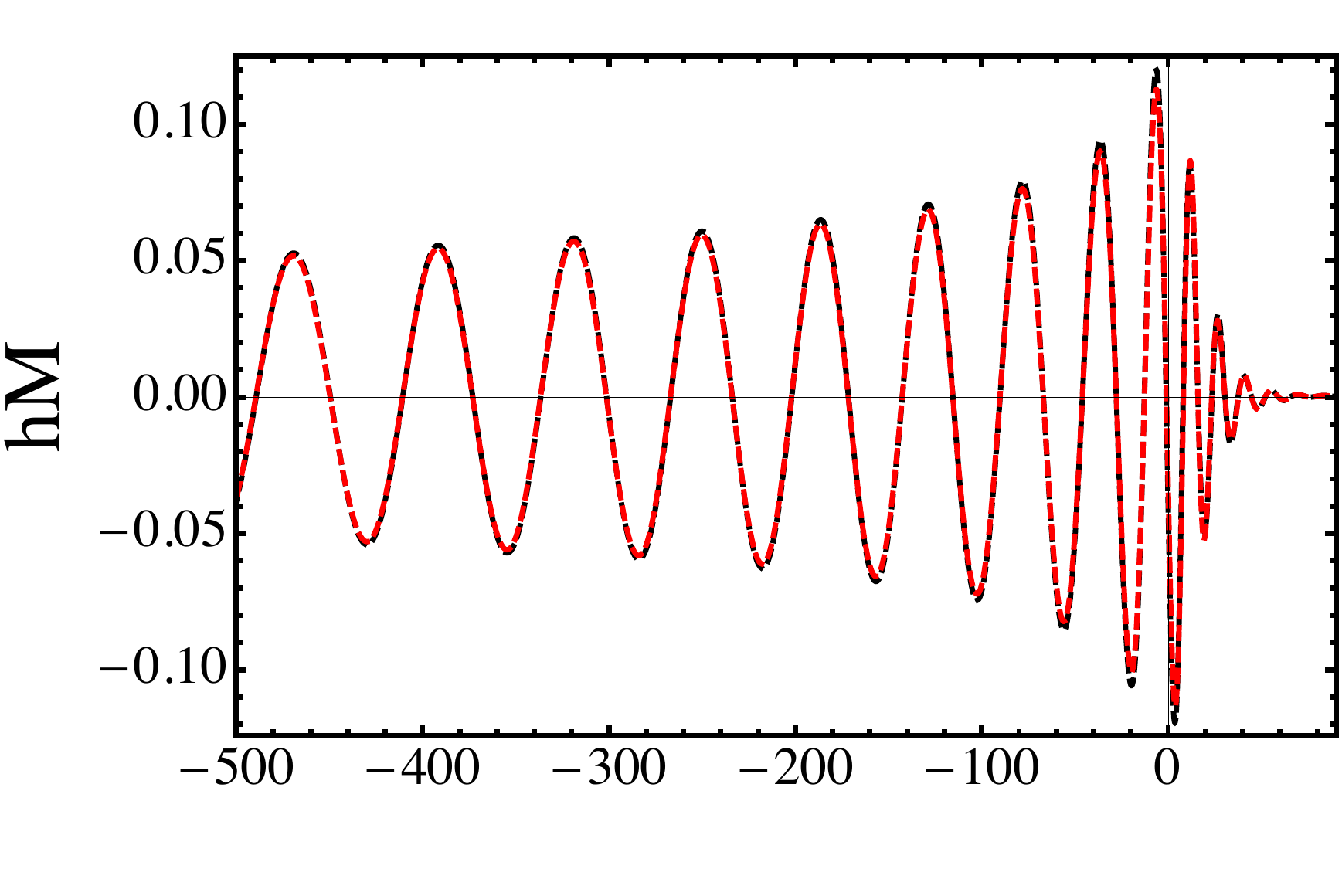}
\includegraphics[width=1.0\columnwidth]{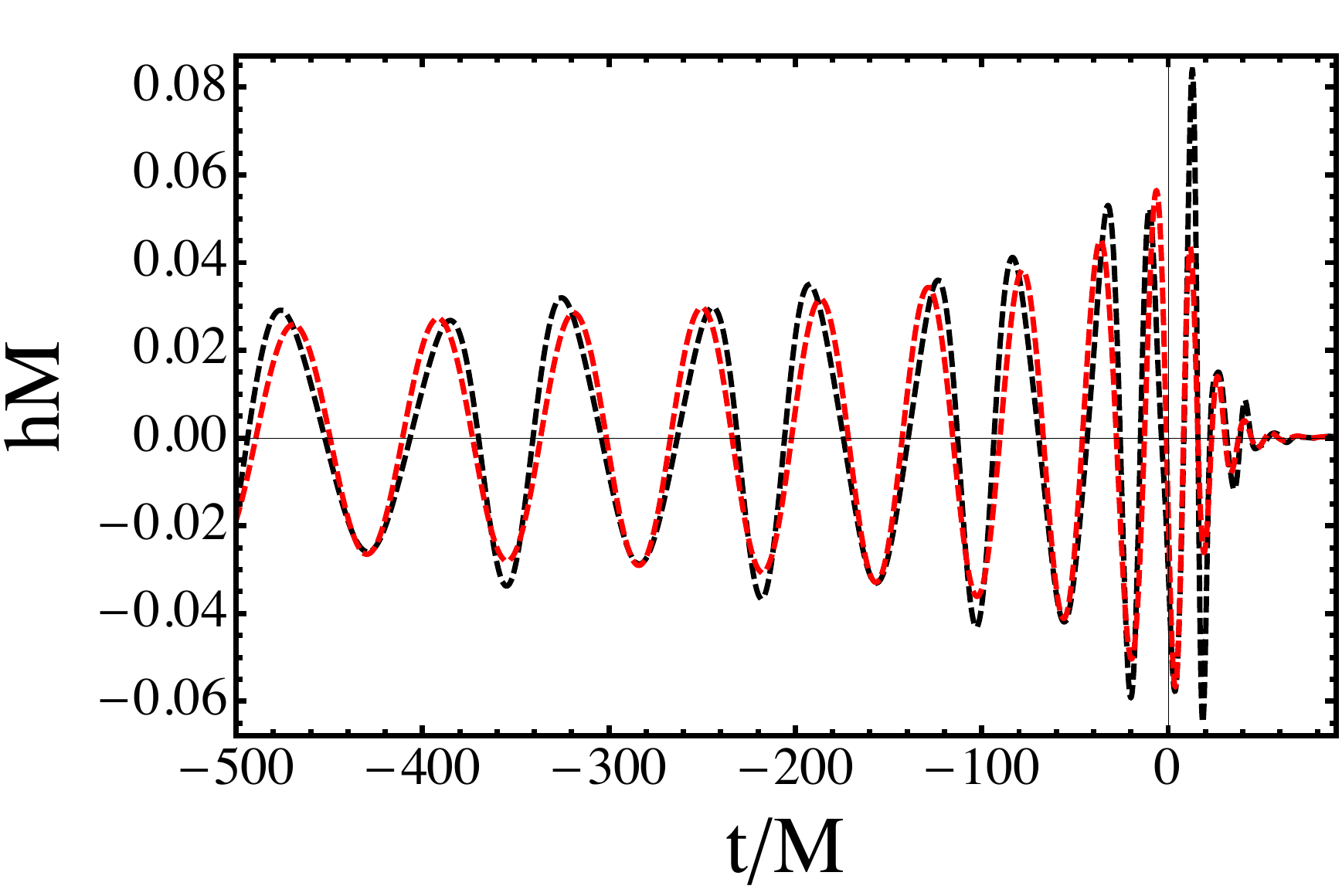}
\caption[Effect of $q$ on higher modes]{\textbf{Impact of higher modes in a time domain waveform:}. We show, in geometrical units, the last few cycles of the full GW signal (black) and only $(2,2)$ mode alone (red) of a mass ratio $q=6$ non-spinning source, when this is face-on (Top panel) and edge-on (Bottom panel). Note that both signals are indistinguishable when the source is face-on, while higher modes cause huge discrepancies in the last cycles of the edge-on case. The simulation used is GT0604 from the Georgia Tech waveform catalogue \cite{Jani:2016wkt}.}
\label{ex:fig:waveforms}
\end{figure}

A BBH is said to be face-on/edge-on to the detector when this is located at $\iota=0$ and $\iota=\pi/2$ respectively. Within this framework, the strain $h$ produced by a GW signal with effective polarization $\psi$ \cite{Bustillo:2015ova} at a given point $p=(d_L,\theta,\varphi)$ on its sky can be decomposed as a sum of modes $h_{\ell,m}(\Xi;t)$ weighted by spin -2 weighted spherical harmonics $Y^{-2}_{\ell,m}(\iota,\varphi)$ as \cite{sharm}:
\begin{equation}
\begin{aligned}
&h(\Xi;d_L,\iota,\varphi;\bar\iota,\bar\varphi,\psi_p;t-t_c)=\\
&F_+(\bar\theta,\bar\varphi,\psi_p) h_+ (d_L;\iota,\varphi;\Xi;t-t_c) + \\ & F_\times (\bar\theta,\bar\varphi,\psi_p) h_\times (d_L;\iota,\varphi;\Xi;t-t_c)=\\
&=\frac{F}{d_L}\bigg{(}\cos\psi {\Re} ({\cal H})+ \sin{\psi}  {\Im} ({\cal H})\bigg{)},
\end{aligned}
\end{equation}
with
\begin{equation}
\begin{aligned}
{\cal H} = h_+ -ih_\times= \sum_{\ell\geq 2}\sum_{m=-\ell}^{m=\ell}Y^{-2}_{\ell,m}(\iota,\varphi)h_{\ell,m}(\Xi;t-t_c).
\end{aligned}
\label{gwmodes}
\end{equation}\\
Above, $F=\sqrt{F_+^2+F_\times^2}$, $\tan \psi=F_\times/F_+$ and $h_{\ell,m}(\Xi;t)=A_{\ell,m}(\Xi;t)e^{-i\phi_{\ell,m}(\Xi;t)}$, $A_{\ell,m}$ and $\phi_{\ell,m}$ being real. As it is common in GW literature, $(h_+,h_\times)$ represent the two polarizations of the GW, the factors $(F_+, F_\times)$ denote the antenna pattern of the detector \cite{Jaranowski:1998qm,Varma:2014jxa} and $d_L$ denotes the luminosity distance.

During most of the inspiral, the $(\ell,m)=(2,\pm 2)$ modes dominate the above sum. The remaining modes, known as higher-order modes, have a sub-dominant effect and only contribute significantly to the GW signal during the last few cycles and merger of the binary. However, the amplitude of the higher-order modes grows with the mass ratio $q$. Also, the spherical harmonics $Y_{2,\pm 2}$ have maxima at $\theta=(0,\pi)$ and minimum at $\theta=\pi/2$, while the rest of the harmonics mostly behave in the opposite way. This makes the $(2,2)$ mode especially dominant for face-on/off sources, while edge-on sources have a more important contribution from higher-order modes \cite{Berti:2007fi,Pekowsky:2012sr,Bustillo:2015qty}.

To a very good accuracy, the frequency $f_{\ell,m}$ of a given $h_{\ell,m}$ mode scales with the orbital frequency of the binary $f_{orb}$ and its total mass $M$ as $f_{\ell,m} \propto m f_{orb}/M$. Hence, as the total mass grows, the $(2,2)$ mode will eventually leave the detector sensitive band, disappearing in the seismic noise wall\footnote{Located at $\sim 25$ Hz for the current version of Advanced LIGO \cite{TheLIGOScientific:2016agk}.}. This exacerbates the contribution of higher-$m$ modes, which have higher-frequency content. This effect is especially important for the case of the current sensitivity curve \cite{TheLIGOScientific:2016agk}, which has its peak sensitivity in the frequency band $100-350$ Hz. The effect of a given mode is generally maximum when its peak amplitude happens at a frequency within this range, as shown for instance in Fig.˜4 of \cite{Bustillo:2015qty}.

\section{Search algorithms}

\subsection{The matched filter algorithm}  
\label{sec:pycbc}

In general, a data segment from a GW detector $s(t)$ will consist on background noise $n(t)$ plus a potential GW signal $g(t)$, with $S_n(f)$
denoting the one-sided power spectral density of $n(t)$. The \pycbc{} pipeline \cite{Canton:2014ena,Usman:2015kfa,pycbc-software} detects the presence of a compact-binary-merger signal by filtering $s(t)$ with a set of template waveforms $\{h\}$, called the \emph{template bank} of the search and covering the target parameter space of compact binary mergers \cite{Allen:2005fk}. Matched filtering leads to a signal-to-noise ratio (SNR) defined as
\begin{equation}
	\rho = \frac{\sqrt{{\langle s | h^I \rangle}^2 + {\langle s | h^Q \rangle}^2}}{\sqrt{\langle h^I | h^I \rangle}}
\end{equation}
with the inner product between waveforms defined as
\begin{equation}
	\langle s | h \rangle = 4 \Re \int_{f_{low}}^{f_{high}}\frac{\tilde{s}(f) \tilde{h}^{*}(f)}{S_n(f)} \mathrm{d}f.
\end{equation} 
Above, $h^{I,Q}$ are two template waveforms differing only by a $90$ degree shift in coalescence phase, $\tilde{h}(f)$ denotes the Fourier transform of $h(t)$ and $^*$ denotes complex conjugation. Matched filtering
produces SNR values at each template for many possible coalescence times covering the available data. Local SNR maxima which cross a predetermined threshold are recorded for each detector as \emph{triggers}.

A source at a given luminosity distance $d_L$ induces an average SNR inversely proportional to $d_L$. The SNR is also proportional to the overlap of the signal $g$ and the best-matching template $h$, defined as \cite{Apostolatos:1995pj}:
\begin{equation}
	(g|h) = \frac{\langle g | h \rangle}{\sqrt{\langle h | h \rangle \langle g | g \rangle}}.
\end{equation}
Thus, if the best-matching template is a poor representation of the signal, the maximum distance at which the source can produce a specific SNR is reduced by a large factor. The omission of higher-order modes from search templates has been shown to exhibit this problem for high-mass and high-mass-ratio systems \cite{Varma:2014jxa,Bustillo:2015qty,Bustillo:2016gid,Capano:2013raa}.

\begin{figure*}
\centering
\includegraphics[width=1.0\columnwidth]{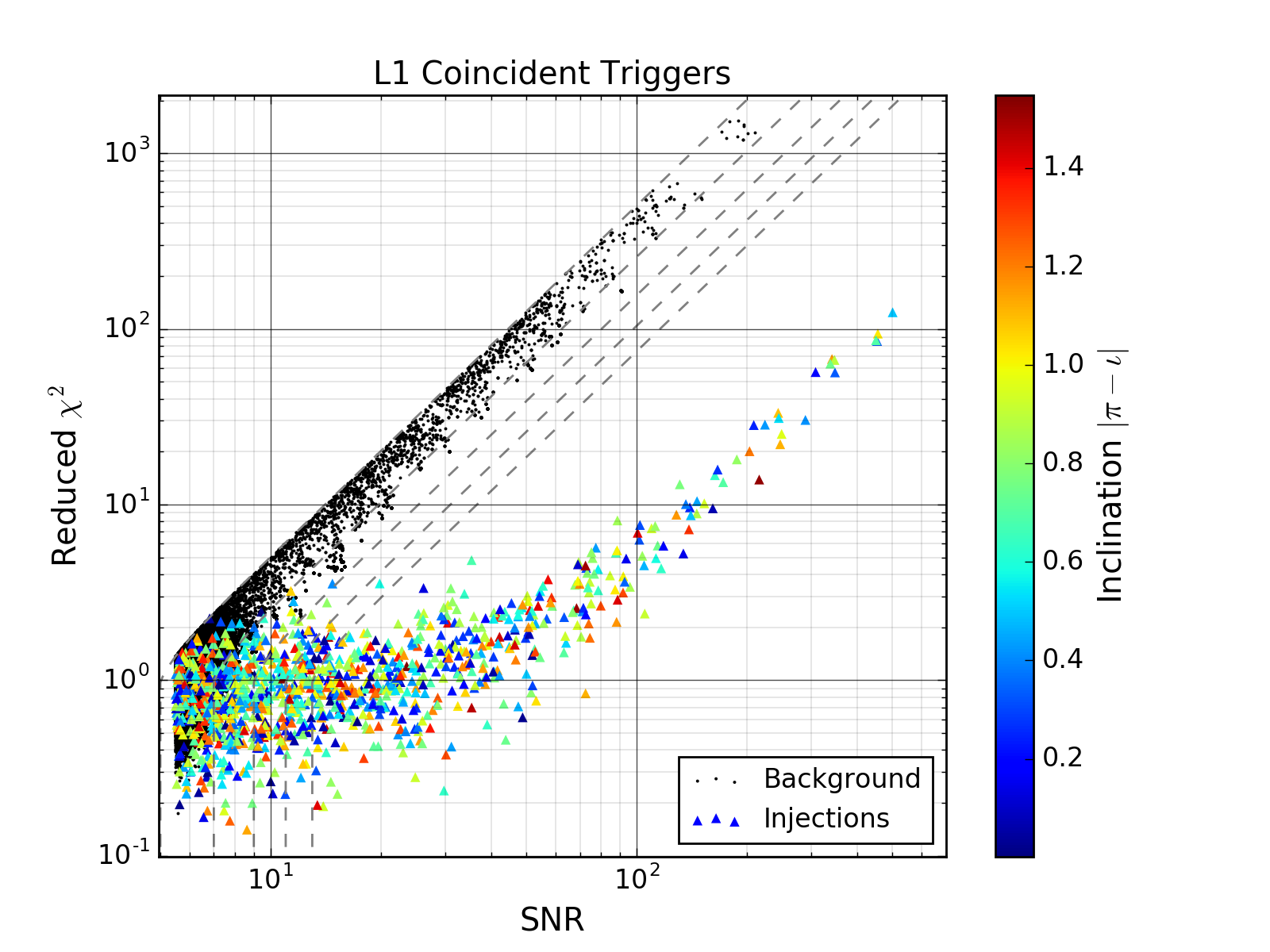}
\includegraphics[width=1.0\columnwidth]{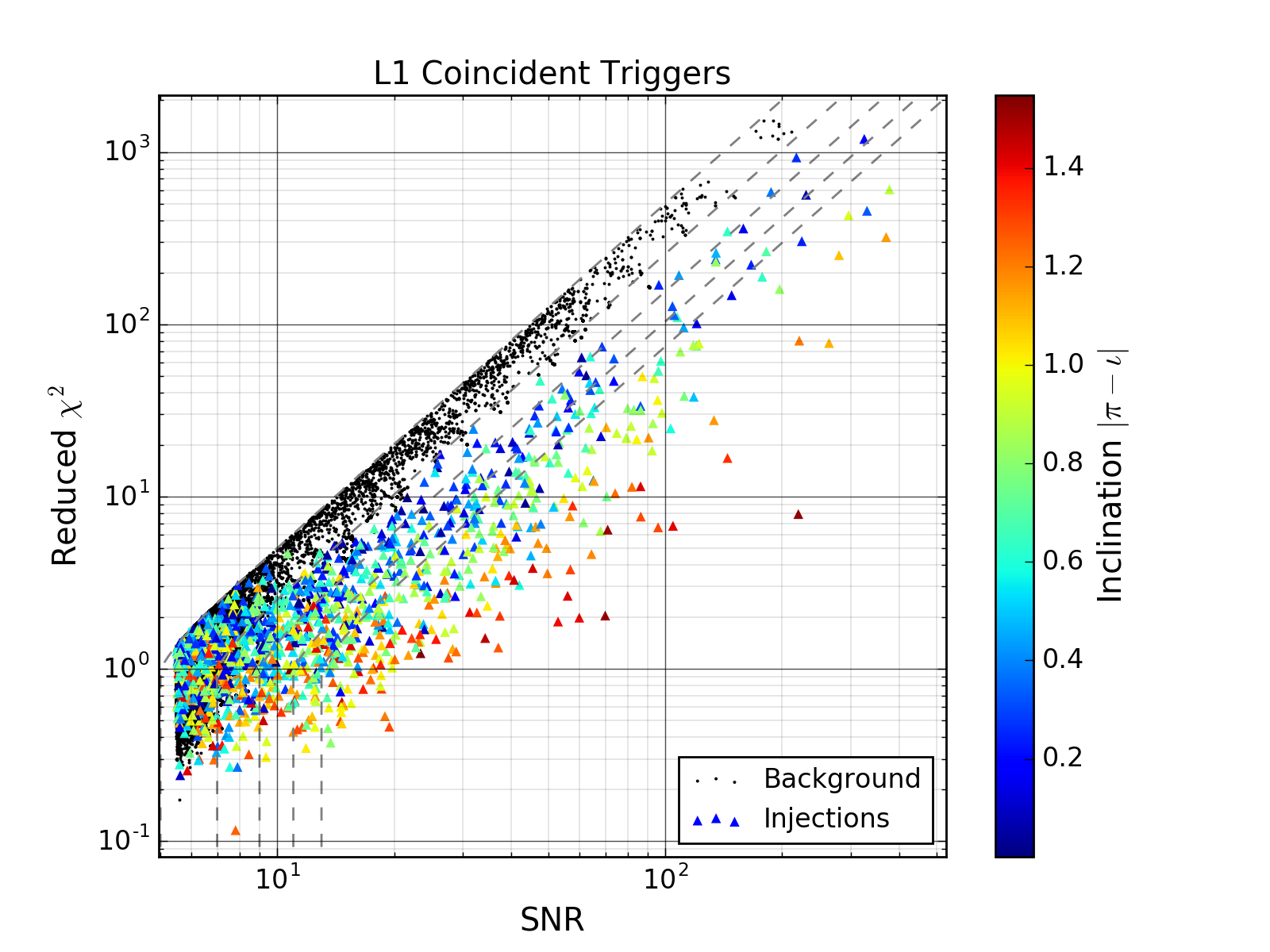}
\caption[Effect of $q$ on higher modes]{\textbf{Impact of higher modes in the \pycbc{} analysis:} The plots show SNR and reduced chi-squared $\rchisq$ for background noise triggers (black dots) and triggers corresponding to simulated signals with $(q=7,M=300M_{\odot})$ (colored dots). The dashed lines denote contours of equal single-detector ranking statistic, roughly representing contours of equal detection significance. The color scale indicates the inclination $\iota$ of the simulated source: blue denotes edge-on sources, red denotes face-on/off ones. In the left panel, simulated signals only consider the quadrupolar radiation mode, while in the right panel several higher-order modes are also included. In the latter case, the mismatch between signal and template raises the value of $\rchisq$, moving the triggers to regions of lower significance, especially
in the case of edge-on sources.}
\label{ex:fig:chisq}
\end{figure*}

Although matched-filtering assumes stationary and Gaussian detector noise, such conditions are not exactly satisfied by real noise in GW detectors. Transient signals of terrestrial origin, known as glitches, often contaminate the data and produce spurious triggers with large SNR \cite{Blackburn:2008ah,Slutsky:2010ff,Canton:2013joa}. The \pycbc{} pipeline uses a signal-based veto or $\chi^2$-test in order to separate astrophysical and spurious triggers \cite{Allen:2004gu}. The time-frequency morphology of the incoming signal $s$ is compared to the template morphology by computing partial SNRs in $p$ disjoint frequency bands, resulting in the reduced $\chi^2$ statistic
\begin{equation}
	\rchisq = \frac{p}{2p-2}\sum_{i=1}^{p}\bigg{(}\frac{\langle s | h_i \rangle}{\sqrt{\langle h_i | h_i \rangle}}-\frac{\rho}{p}\bigg{)}^2,
\end{equation}
Above, $\rho$ is the SNR of the trigger and $h_i$ is the template $h$ restricted to the $i$-th frequency band. The bands are chosen to contain an equal fraction of the total SNR if the template is a correct representation of the data. When this happens, or when the data contains only stationary Gaussian noise, $\rchisq$ is close to unity; in the presence of a loud signal whose morphology is inconsistent with the template, $\rchisq \gg 1$.

Single-detector triggers are then ranked by a function of SNR and $\rchisq$ known as reweighted SNR $\hat\rho(\rho,\rchisq)$, which basically down-weights triggers with large $\rchisq$. Triggers in suitable coincidence between the Hanford and Livingston interferometers are assigned a network rank based on the individual reweighted SNRs. The background distribution of the network ranking statistic is obtained empirically by repeating the coincidence many times, each using a different unphysical time shift between the two detectors. Each foreground (unshifted) coincidence is then compared to the background distribution and assigned a statistical significance, corresponding to the (inverse) rate of false coincidences with a higher network rank. We will refer to this significance as inverse false alarm rate (IFAR). For more details on the procedure, see \cite{Usman:2015kfa}.

The \pycbc{} search configuration considered here corresponds to what was used for the most recent IMBHB rate upper limits by the LIGO and Virgo collaborations \cite{Abbott:2017iws}. It uses template waveforms computed via a phenomenological, non-precessing waveform model considering only the dominant $(2,2)$ mode of the radiation field \cite{Husa:2015iqa,Khan:2015jqa}\footnote{The \phenomd{} approximant from the LALSimulation library \cite{LALSimulation}.}. The template bank covers systems with total mass $M\in[50,600]M_{\odot}ç$, mass ratio $q\in[1,10]$ and dimensionless spin projections along the orbital angular momentum $\chi_{1,2}\in [-0.99,+0.99]$, where positive values of the spin indicate prograde spin with respect to the orbit. The lower frequency cutoff for matched filtering is set to $f_{low}=20$ Hz.

\subsection{The un-modeled algorithm}

Unlike the matched-filter search, the un-modeled Coherent Waveburst algorithm $\verb+cWB+$ does not rely on any a-priori knowledge of the morphology of the incoming GW signal. Due to this, it can be used for searching for a broader range of signals, like those coming from eccentric binary black holes \cite{Tiwari:2015gal}.  This algorithm decomposes the incoming data in time-frequency maps. Once this is done, coherent triggers coming from the same region of the time-frequency maps having a power that exceeds that of the one expected from noise are considered as candidate triggers.
Next, candidate triggers are reconstructed 
by means of a constrained maximum-likelihood analysis.
Because this reconstruction does not make any prior assumption about the shape of the signal, $\verb+cWB+$ might in principle be used to search for those sources that the matched-filter search fails to recover. The obvious example here is that of IMBHB signals with strong higher-modes content. Future follow-up studies will also focus on precessing BBH.\\
In this work we use the same pipeline configuration used in \cite{Abbott:2017iws}. As described there, we apply a constraint in the morphology of the GW signals, in order
to favor the reconstruction of chirality-polarized
waveforms \cite{Klimenko:2015ypf}. Also, modifying the set-up considered in \cite{Abbott:2016ezn}, frequency-varying post-production
selection cuts are tweaked to minimize
their impact on the sensitivity to IMBHB mergers: low frequencies are likely to be
contaminated by several environmental and instrumental
artifacts that can mimic waveforms for
massive binary mergers. A well known example of these are the so called sine-Gaussian glitches \cite{Canton:2013joa}. As in \cite{Abbott:2017iws} we start the $\verb+cWB+$ analysis at a frequency of $f_{low}=16$Hz.
For a thorough description of the $\verb+cWB+$ algorithm, see \cite{Klimenko:2008fu,Klimenko:2015ypf}.

\section{Estimating the sensitivity of a search}
\subsection{MonteCarlo simulations}

The sensitivity of a search algorithm to a given system can be estimated by simulating the corresponding signal, adding it to realistic data, analyzing the data
with the search algorithm and recording whether the signal has been detected at the given significance, usually expressed by the IFAR.

Consider a set of $N_{tot}$ injections distributed uniformly in a co-moving volume $VT_{tot}[$Gpc$^3$ yr$]$. The sensitive volume of the search at a given reference IFAR $x_0$ can then be evaluated by computing the number of injections $N_{rec}$ recovered with an IFAR $x > x_0$ as:
\begin{equation}
\langle VT \rangle = \frac{N_{rec}}{N_{tot}} \langle VT_{tot} \rangle
\end{equation}
In our case, we distribute our sources uniformly in co-moving volume up to luminosity distances corresponding to redshifts of $z=0.3$ and $z=0.8$, depending on the intrinsic parameters of the source. The injections cover a total period of $130$ days, which leads to respective total surveyed co-moving volumes of  $\langle VT_{tot} \rangle = 2.6$ and $22.26$ Gpc$^3$ yr. The orientation of the sources is uniformly distributed in $\varphi$ and $\cos(\theta)$.

\begin{figure*}
\centering
\includegraphics[width=0.95\columnwidth]{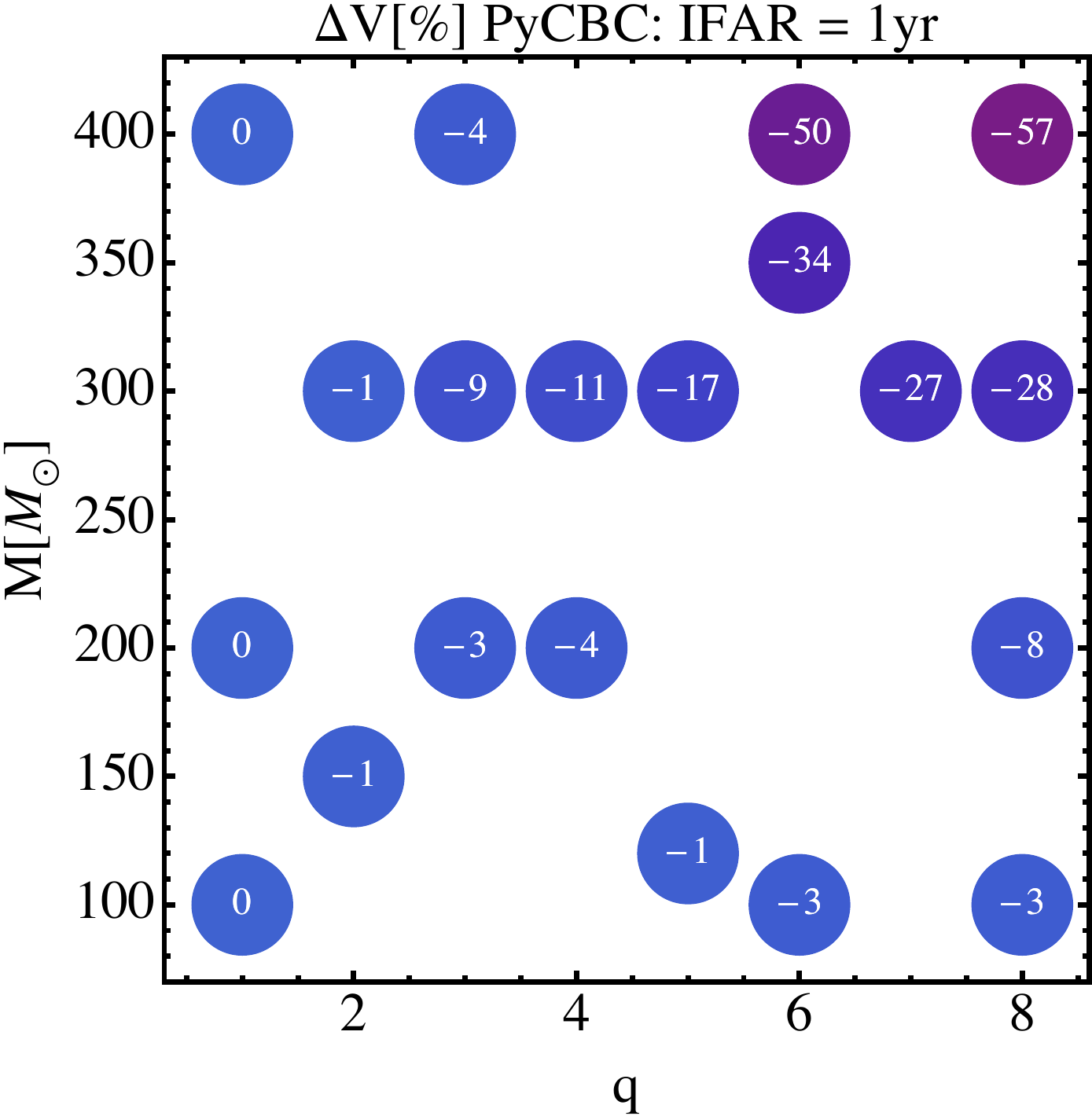}
\includegraphics[width=0.95\columnwidth]{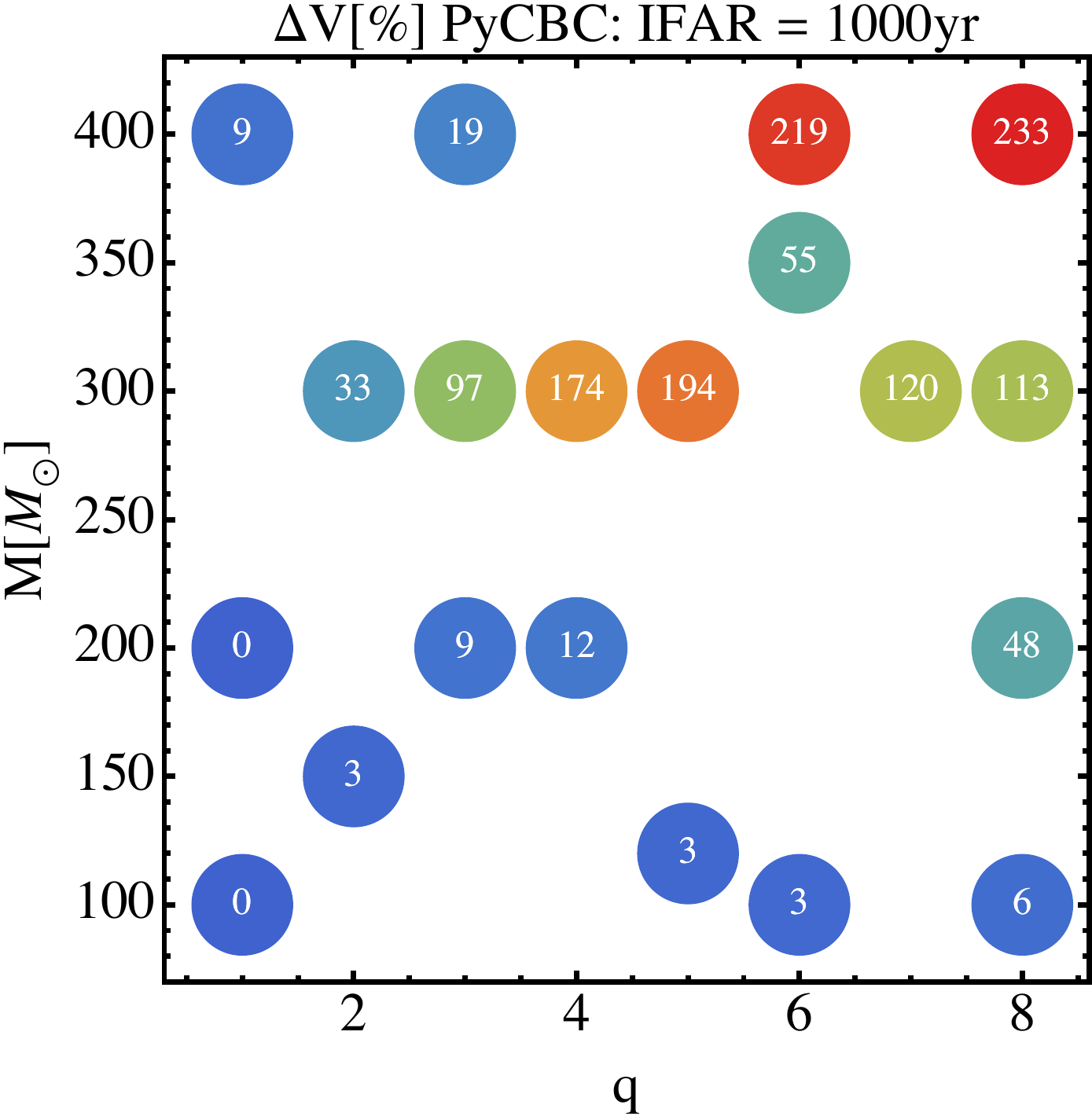}
\caption[Effect of $q$ on higher modes]{\textbf{\pycbc{}: $(2,2)$ mode vs. higher-order modes:} $\Delta V$ values for \pycbc{}. The left plot shows results for IFAR$=1$yr while the right one corresponds to IFAR$=10^3$yr. Values close to $0$ denote a negligible impact. At low IFAR, the additional signal power in the higher modes makes the search more sensitive, leading to negative $\Delta V$ values. At large IFAR, the dominant effect is the worsening of the $\rchisq$ due to the mismatch between the signals and the quadrupolar templates, producing the opposite effect. In particular, values of $\Delta V = 200\%$ are reached in the large-mass-ratio, large-mass limit.}
\label{ex:fig:pycbc}
\end{figure*}

The intrinsic parameters of the simulated sources are those shown in Fig.\ref{ex:fig:pycbc}. In this study, total masses are quoted in the source reference frame. 
For each choice of mass parameters, we prepare two sets of injections with identical parameters which are then injected in the O1 LIGO Science Run data. The injection sets only differ in their omission or inclusion of higher-order modes, as described below.

We then run both the \pycbc{} and $\verb+cWB+$ pipelines on these injection sets and estimate the corresponding comoving sensitive volumes $ \langle VT \rangle_{Nohigher-modes} $ and $ \langle VT \rangle_{higher-modes}$ at a fixed IFAR. From these, we compute the relative fractional overestimation of the sensitive volume, produced by the omission of higher modes as:
\begin{equation}
\Delta V [\%] = 100 \times \bigg{(} \frac{\langle VT\rangle_{Nohigher-modes}}{\langle VT\rangle_{higher-modes}} - 1\bigg{)}
\end{equation}
Positive values indicate that omitting higher-order modes causes an overestimation of the sensitivity of the search while negative ones indicate the opposite. As we will show, the sign and magnitude of $\Delta V$ will depend not only on the properties of the target source but also on the threshold IFAR required and the search algorithm considered.

\subsection{Waveform models}

We simulate our signals as non-spinning waveforms computed within the Effective One Body framework (EOB). We consider two waveform models that respectively omit and include higher-order modes, but are otherwise very similar: $\verb+EOBNRv2+$ and $\verb+EOBNRv2HM+$, both available in the LALSimulation library \cite{LALSimulation}. $\verb+EOBNRv2HM+$ includes the $(\ell,m)=(2,\pm1),(2,\pm2),(3,\pm3),(4,\pm4)$ and $(5,\pm5)$ modes \cite{Pan:2011gk}. Our choice is based on the fact that no other models considering higher-modes were available at the time of this study. Moreover, the upper limits on IMBHB coalescence published by the LIGO and Virgo collaborations mostly considered non-spinning sources \cite{Abbott:2017iws}, which makes our target signals suitable for comparison with their results. Note that aligned-spin models for mergers containing higher-order modes are now making fast progress \cite{London:2017bcn,COTESTA}. Regarding our choice of a model omitting higher-modes, we could have chosen from several other waveform models, for instance the phenomenological model \phenomd{} \cite{Husa:2015iqa,Khan:2015jqa} or the effective-one-body model $\verb+SEOBNRv2+$ \cite{Taracchini:2012ig,Taracchini:2013rva, Buonanno2013}. However, we want to make sure that any difference we observe in the results including and omitting higher-modes are indeed due to higher modes, not to differences between the modeling of the $(2,2)$ mode.

\begin{figure*}
\centering
\includegraphics[width=.95\columnwidth]{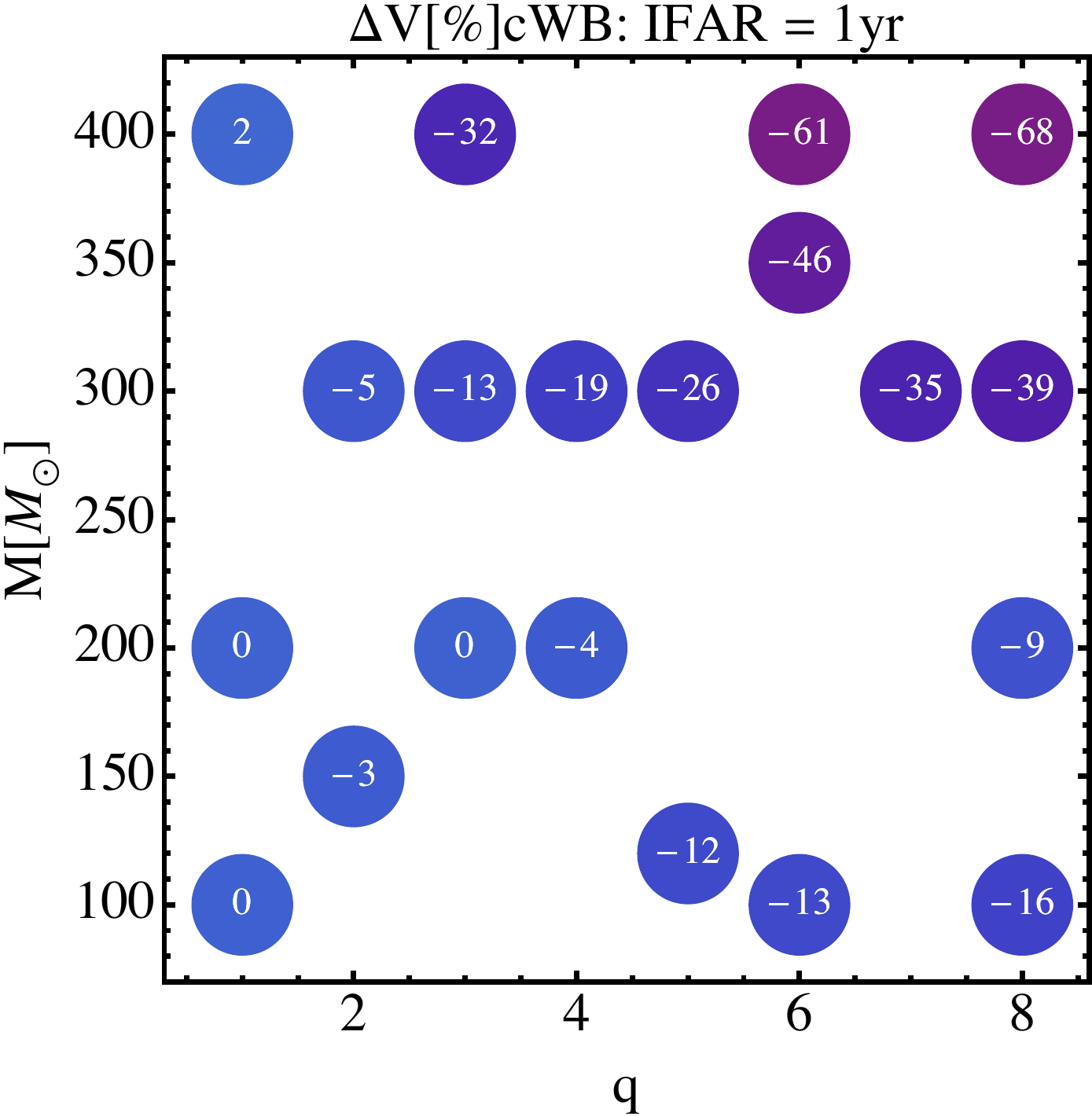}
\includegraphics[width=.95\columnwidth]{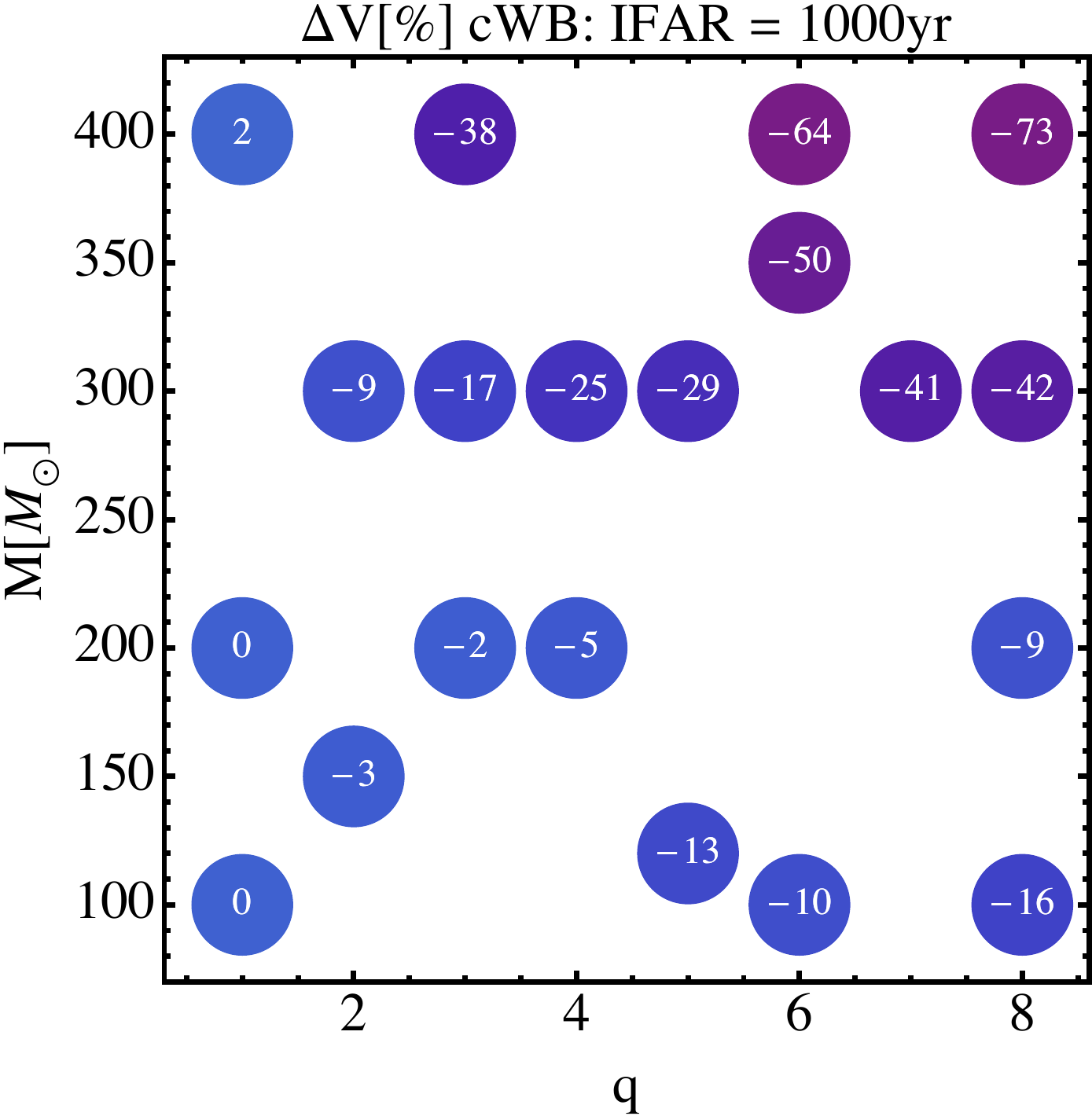}
\caption[Effect of $q$ on higher modes]{\textbf{cWB: $(2,2)$ mode vs. higher-order modes:} $\Delta V$ values for cWB. The left plot shows results for IFAR$=1$yr while the right one corresponds to IFAR$=10^3$yr. Values close to $0$ denote a negligible impact. Note that for almost all cases, and unlike for \pycbc{}, the increment of the GW signal power due to the inclusion of higher modes enhances the sensitivity of cWB, leading to negative $\Delta V$ values for all IFARs. Again, this effect is larger in the large mass ratio - large mass limit.}
\label{ex:fig:cWB}
\end{figure*}

\section{Results}

In a recent study, the LIGO and Virgo collaborations reported upper limits on IMHBH merger rates, estimated after analyzing the data taken in Advanced LIGO's first observing run \cite{Abbott:2017iws}. The upper limits were obtained by combining the sensitivities of two search algorithms to simulated signals: the matched-filter pipeline known as $\verb+gstlal+$ \cite{Cannon:2011vi,Cannon:2011xk,Cannon:2012zt} and the un-modeled pipeline $\verb+cWB+$. Another matched-filter search pipeline, \pycbc{}, was used for cross-checking the $\verb+gstlal+$ analysis, finding consistent results. The simulated signals omitted the contribution of higher-order modes and, as previously mentioned, this can lead to biased sensitivity and rate estimates.

In this section we compute and compare the sensitivities of the $\verb+cWB+$ and \pycbc{} analyses, configured as in \cite{Abbott:2017iws}, to simulated signals which omit and include higher-order modes. We estimate the biases produced by the omission of higher-order modes in the upper limits presented in \cite{Abbott:2017iws}. We also compare the sensitivity of the matched-filter and unmodeled analyses across the considered parameter space.

\subsection{Matched-filter search}

As described in subsection IIIA, the matched-filter search considered here ranks single-detector triggers using the statistic $\hat\rho(\rho,\chi^2)$.   Large SNR and low $\rchisq$ values lead to larger values of $\hat\rho$, which typically correspond to larger detection significance.

Fig.~\ref{ex:fig:chisq} shows the values of SNR and $\rchisq$ obtained for two injections sets and the corresponding background triggers. The injections in the two sets differ in the choice of waveform model, but have otherwise identical parameters. In particular, they correspond to $q=7$ binaries with a total mass of $300M_\odot$ with varying orientations and sky locations. The color code denotes the value of the inclination $\iota$. Red denotes face-on/off sources and blue denotes edge-on ones. Finally, the gray dashed lines are contours of constant reweighted SNR, which approximate contours of constant detection significance.\\
In the left panel, the injections contain only the dominant $(2,\pm2)$ modes. In this case, the search templates \phenomd{}, which also consider only the $(2,\pm2)$ mode, are good representations of the injected signal. All injections are well separated from the background, for large enough SNR. The $\rchisq$ values are also independent from the inclination of the source.\\
The right panel shows the same quantities, but in this case the injections contain higher-order modes. Note how for a given SNR, edge-on injections show larger $\rchisq$ values than face-on/off ones, causing triggers to move to lower-significance regions and mix with background triggers. This is due to the fact that face-on signals are vastly dominated by the $(2,2)$ mode while the higher-order modes become more important as the inclination of the system grows, making the injected signal differ from the template that recovers it. Moreover, adding higher-order modes to the injected signals increases the amount of signal power in the sensitive band of the interferometers. Due to occasional overlap between the higher modes and the quadrupolar templates, the additional power can in some cases boost the recovered SNR.  As we will see, the impact of these two effects on the sensitivity of the search depends on the minimum IFAR required for a detection. A larger IFAR requires a better separation of the injected signals from the background noise, which is not possible when higher-order modes are strong.

\begin{figure*}
\centering
\includegraphics[width=.95\columnwidth]{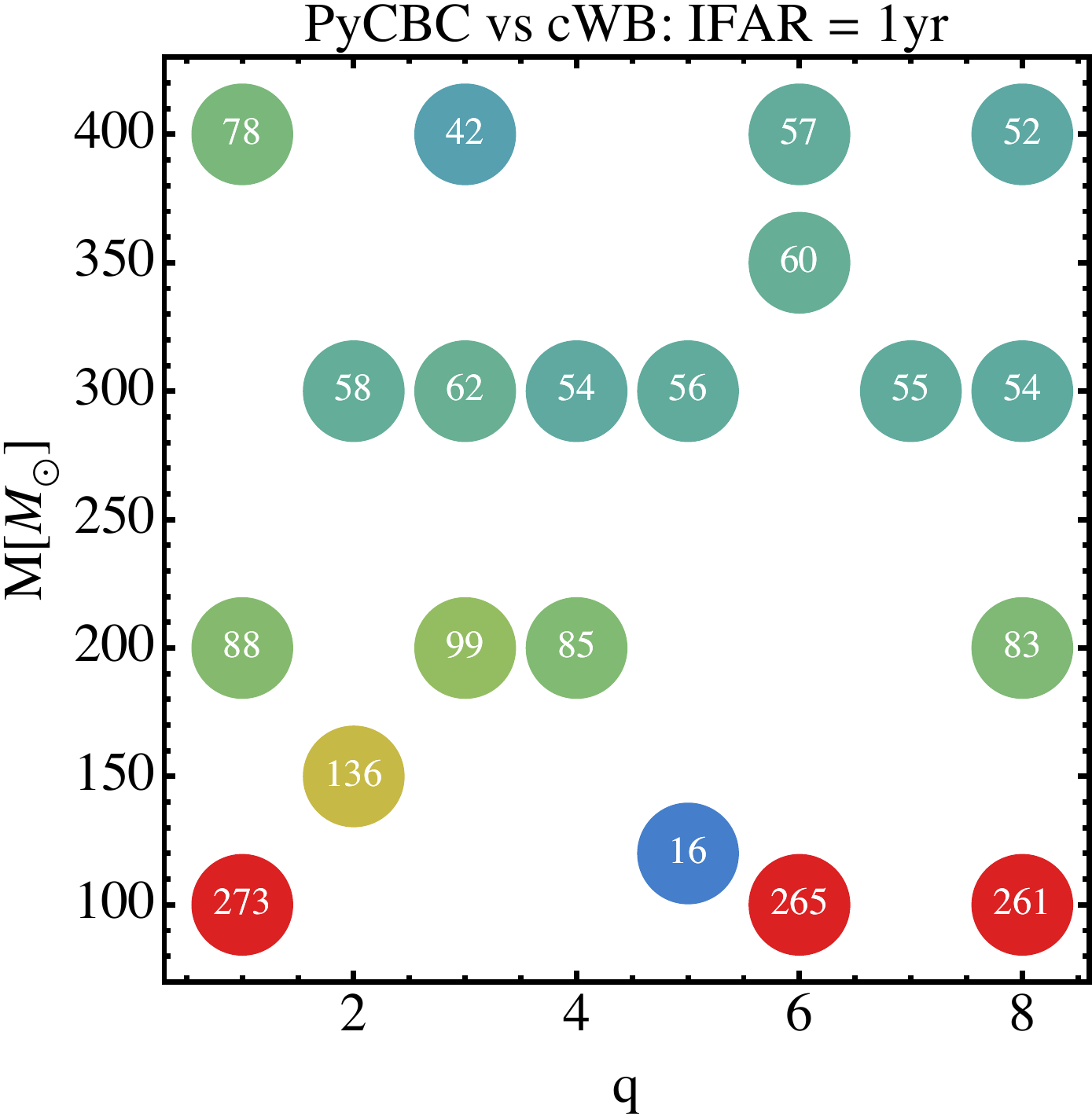}
\includegraphics[width=.95\columnwidth]{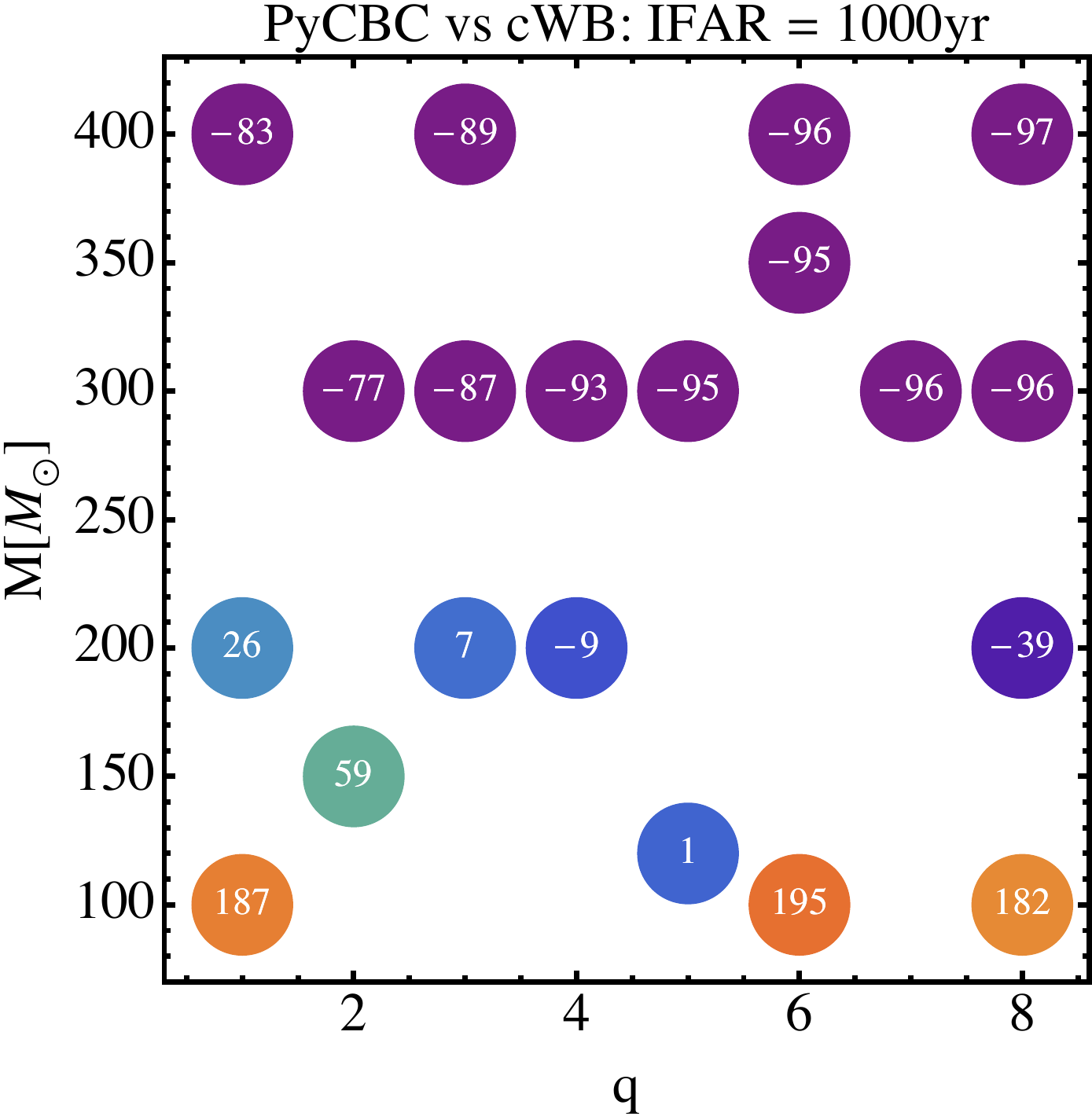}
\caption[Effect of $q$ on higher modes]{ \textbf{\pycbc{} vs. CWB including higher modes:} Comparison of sensitive volumes of cWB and \pycbc{} when higher modes are included. Here we take $\Delta V = 100(V_{PyCBC}/V_{cWB} -1)$. At low IFAR, the \pycbc{} search always outperforms cWB. The effect of higher modes on \pycbc{} and cWB causes cWB to outperform \pycbc{} by a larger margin than in the case in which higher modes are omitted (see Fig.6). Errors on the numbers are typically $\pm 2\%$ and at most $\pm 6\%$.}
\label{ex:fig:cWBpyhigher-modes}
\end{figure*}

\begin{figure*}
\centering
\includegraphics[width=.95\columnwidth]{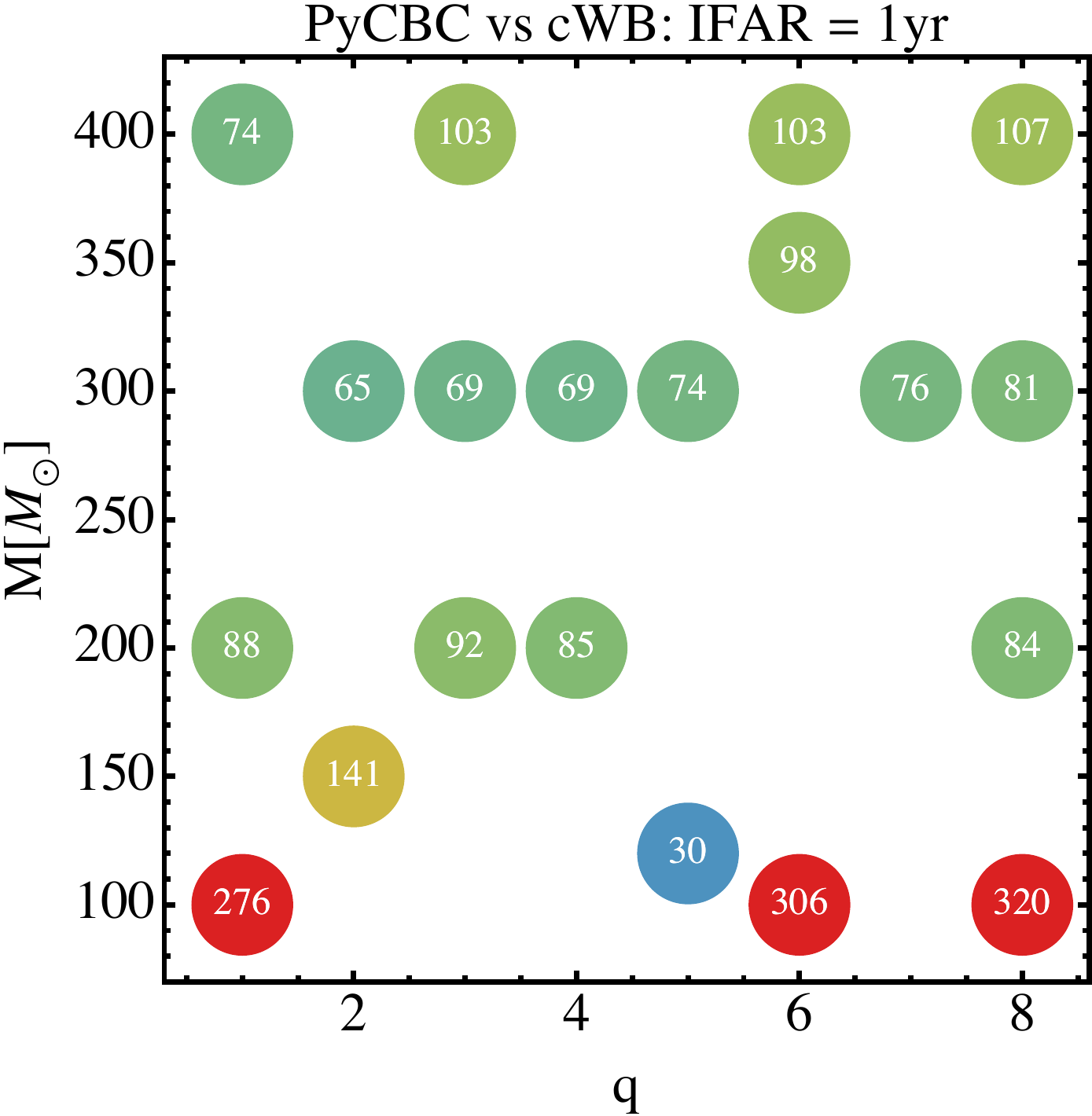}
\includegraphics[width=.95\columnwidth]{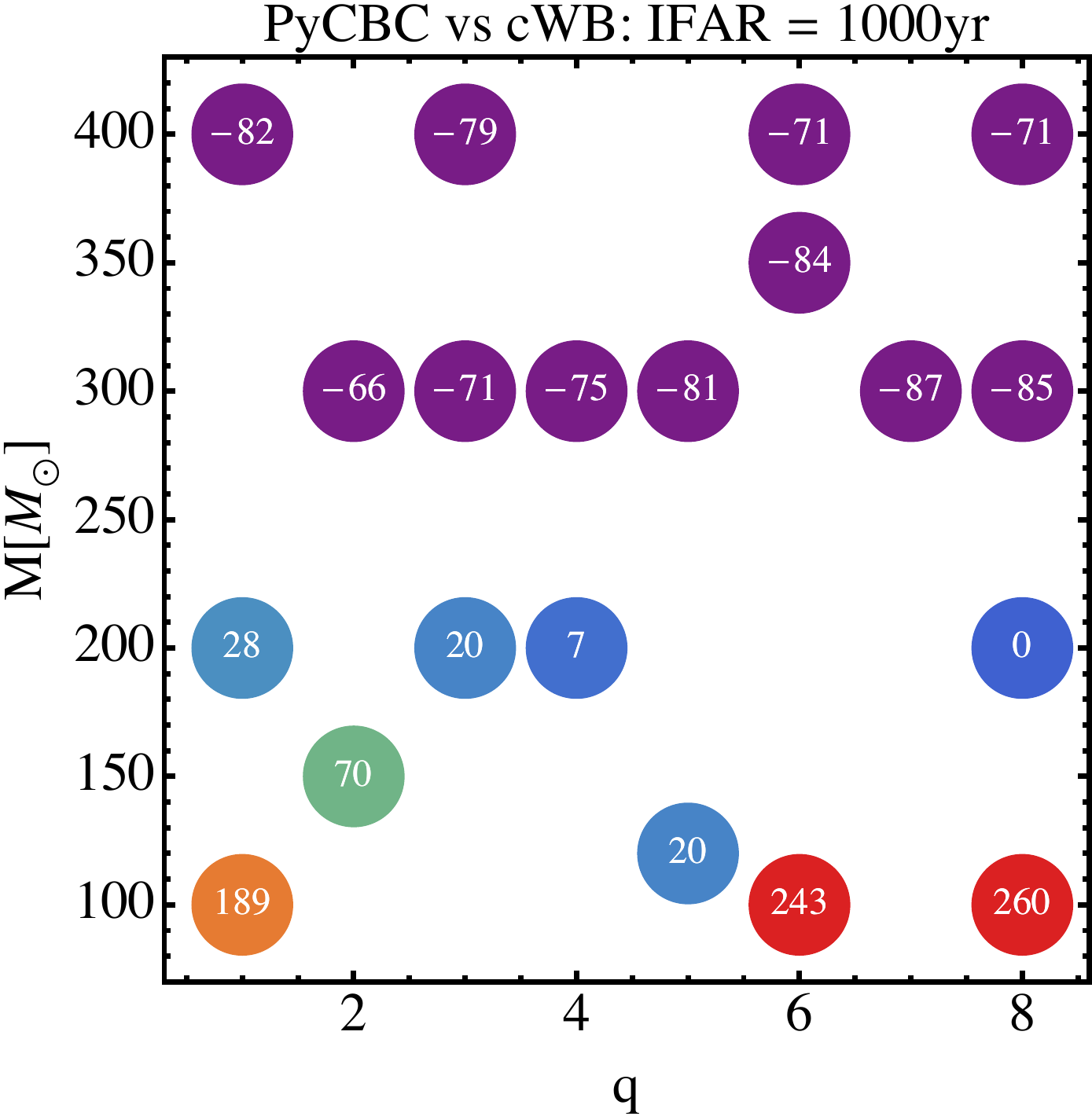}
\caption[Effect of $q$ on higher modes]{\textbf{\pycbc{} vs. CWB omitting higher modes:} Comparison of sensitive volumes of cWB and \pycbc{} when higher modes are omitted. Here we take $\Delta V = 100(V_{PyCBC}/V_{cWB} -1)$. At low IFAR, the \pycbc{} search always outperforms cWB. At large IFAR, cWB is always more sensitive than PyCBC to sources with $M > 200M_\odot$ Errors on the numbers are typically $\pm 2\%$ and at most $\pm 6\%$.}
\label{ex:fig:cWBpy22}
\end{figure*}

Fig.\ref{ex:fig:pycbc} shows the percent difference $\Delta V [\%]$ of the sensitive volumes obtained, for a given inverse false alarm rate (IFAR) for two different IFAR thresholds. Typical errors for the numbers we show in all plots are of $\pm 2\%$ while maximum ones reach $\pm 6\%$. Negative values indicate that the sensitivity of the search is underestimated when higher-order modes are omitted while positive ones indicate the opposite. For the largest IFAR considered ($10^3$yr) and highly asymmetric binaries, the overestimation of the sensitive volume can reach values $ > 200\%$ due to the significant contribution of higher-order modes to the signal. At lower total mass and mass ratio, instead, higher-order modes do not appear to affect the total signal sufficiently, as $\Delta V [\%]$ is consistent with zero for such sources.

Recently, the LIGO LSC published upper limits on the rate of IMBHB coalescences using a reference IFAR of $\sim 0.4$ years \cite{Abbott:2017iws}. For this reason, we repeat our evaluation of the impact of higher-order modes at a comparable IFAR of 1 year. Unlike the previous case, we now find that omission of higher-order modes actually causes an underestimation of the sensitivity. With such a low IFAR threshold, a strict separation of injections from the background is not required. The extra signal power due to the higher-order modes, which can be partially recovered by the \phenomd{} templates, has a bigger effect than the worsening of the $\chi^2$ test, leading to a larger number of detections.

\subsection{The un-modeled search}

While the matched-filter pipeline requires a good match between the incoming GW signal and one of the templates, the un-modeled pipeline remains agnostic about the morphology of the signal. For this reason it is expected that higher modes should improve the sensitivity of $\verb+cWB+$ in all cases. Fig.~\ref{ex:fig:cWB} shows the same results presented in the previous section, for the case of $\verb+cWB+$. Unlike \pycbc{}, the only effect higher-modes have here is to increase the signal power in the sensitive band of the interferometer, boosting $\verb+cWB+$'s ability to pick up the signal and thus the significance of the triggers. This effect is more visible for high mass-ratio and high mass sources. At low mass, the impact of higher-modes in the signal is negligible and, as for the case of \pycbc{}, there is no difference between the sensitive volumes obtained when omitting and including them. On the other hand, for large masses and mass-ratios, the higher-mode content of the signal increases significantly the power that $\verb+cWB+$ can recover, leading to important improvements of the sensitive volume, indicated by negative values of $\Delta V [\%]$. In the most extreme cases, the higher-modes cause an increase of more than $230\%$ of the sensitive volume obtained when higher-modes are omitted, which corresponds to our reported $\Delta V = -71\%$ for the $(q,M)=(8,400)$ case.

As we will see later, for large masses and at an IFAR of $1000$yr, which can be taken as a lower threshold for a confident detection, the sensitivity of the un-modeled algorithm is much larger than that of the matched-filter one. On top of this, as we have just seen, the sensitivity of the former is largely increased when higher-modes are considered. This indicates that BBH coalescence rate studies would be largely overestimated by an omission of higher modes. We will discuss this in a later section.

\subsection{Comparing the matched-filter and un-modeled searches}

In this subsection we compare the sensitivity of both pipelines to the same set of injections. It is known that in stationary Gaussian noise, the matched-filter is the optimal way for finding signals of known morphology, assuming the templates are good representations of the sought-after signals. As we have just seen, this assumption is not true when the templates omit the higher-order modes of the GW signals. For this reason, the un-modeled search might perform better than matched-filtering when higher modes are included in the incoming signals, especially for sources for which higher-modes give a non-negligible contribution.

Fig.~\ref{ex:fig:cWBpyhigher-modes} shows the percent differece of the sensitive volumes of \pycbc{} and $\verb+cWB+$ to our injection sets. At low IFAR (left panel), a good separation of injections from background is not required, and \pycbc{} outperforms $\verb+cWB+$ in all regions of the parameter space. At large IFAR, the sensitivity of \pycbc{} is significantly hindered by the presence of higher modes while that of $\verb+cWB+$ is enhanced. At low mass, this makes \pycbc{} outperform $\verb+cWB+$ by lower margins than at low IFAR. At large mass, we see that $\verb+cWB+$ vastly outperforms \pycbc{}. $\Delta V$ values get as low as $\Delta V = -97\%$, meaning that the un-modeled search is an order of magnitude more sensitive than the matched-filter one.

Assuming that noise is stationary and Gaussian, one might think that the larger sensitivity of the un-modeled search at large mass and large IFAR is solely due to the impact of higher-order modes. However, real detector noise is only approximately stationary and Gaussian. This decreases importantly the performance of the matched filter on its own, requiring the usage of further data analysis techniques like the $\chi^2$ veto (see section IV). The effectiveness of such techniques varies strongly across the search parameter space, however. In particular, many short noise transients turn out to be quite similar to some high-mass binary-merger signals, making signal-based vetoes like the $\chi^2$ test much less effective at rejecting them.

In order to demonstrate such effect further, Fig.~\ref{ex:fig:cWBpy22} shows the same as Fig. \ref{ex:fig:cWBpyhigher-modes} for the case when higher-modes are omitted from the signals. The results are quantitatively very similar to those obtained when higher modes are included. In particular, at large mass and large IFAR, we obtain values of up to $\Delta V = -82\%$, i.e.~the un-modeled search can be 5.5 times more sensitive than the matched-filter one at large masses, even though the matched-filter templates are good representations of the injected signals. This suggests that $\verb+cWB+$ has a greater ability to reject short noise transients than \pycbc{}. In fact, signal-noise discriminators like the $\chi^2$ veto implemented by \pycbc{} are known to have poor performances for short duration signals.

\section{Impact on coalescence rates}

In \cite{Abbott:2017iws}, the LIGO LSC computed results for upper limits of IMBHB coalescence inferred from the LIGO O1 Science Run data. In doing it, the impact of higher-order modes in the model of the real GW was omitted. 
Rates were obtained after evaluating the sensitivity of a joint search performed by the un-modeled pipeline $\verb+cWB+$ that we have used, and the matched-filter pipeline $\verb+gstlal+$. The \pycbc{} pipeline was also run for verification and showed results consistent with $\verb+gstlal+$. Since, as we have shown, omission of higher-modes can bias the sensitivity of GW search pipelines, it is immediate to ask what the impact on the corresponding inferred rates would be. In short, an overestimation of the sensitivity of the search would lead to a underestimation of the corresponding rates, and viceversa.

We have demonstrated that for an IFAR of $1000$yr, omission of higher-modes can cause a very strong bias in the inferred sensitivities and corresponding rates. For total masses $M<200M_{\odot}$, the matched-filer pipeline is more sensitive than the un-modeled one even when higher-modes are included. Hence, its sensitivity will dominate that of the joint search. Since omission of higher-modes leads to an overestimation of this sensitivity, the corresponding upper limits would be underestimated for large mass ratio sources. The converse would happen at larger masses, for which the sensitivity of the joint search would be dominated by that of the un-modeled search, for which the sensitivity is underestimated by omitting higher-order modes.

It is crucial to note however, that the results quoted in \cite{Abbott:2017iws} used a reference IFAR of less than $1$yr. At this IFAR, the sensitivity of the joint search is dominated by that of the matched-filter algorithm, as can be noted in the left panel of Fig.8 and in Table I of \cite{Abbott:2017iws} and in the left panels of Figs 5 and 6. At such low IFAR, we have shown that omission of higher-order modes actually underestimates the sensitivity of the matched-filter algorithm, leading to an overestimation of the corresponding upper limits for the case of highly asymmetric binaries. In other words, whenever higher-modes are important, their omission leads to conservative estimates of the IMBHB coalescence upper limits. This means that the corresponding upper limits quoted in \cite{Abbott:2017iws} are conservative estimates of the ones that would be obtained if higher-order modes were taken into account. In addition, the most stringent upper limits quoted in \cite{Abbott:2017iws}, obtained for nearly equal mass sources, would remain unchanged due to the negligible impact of higher-order modes in the corresponding GW emission.

\section{Conclusions}

The GW emission from IMBHB can have a strong higher-order modes content. This has been traditionally omitted by studies evaluating the sensitivity of GW searches for these objects. In this work we evaluated the impact of this omission on two pipelines used on LIGO O1 Science RUN data to search for IMBHB: an un-modeled search ($\verb+cWB+$), and a matched-filter search (\pycbc{}). To this end, we compared the sensitive volumes $\langle VT \rangle$ of both pipelines estimated with sets of simulated signals both omitting and including higher modes.

The results show that for all significance thresholds considered, the sensitivity of $\verb+cWB+$ is underestimated due to the omission of the power contained in the higher-order modes, when the studied sources have a large total mass $M$ and mass ratio $q$. This is also true for the case of \pycbc{}, but only in the low significance limit, in which a clear separation between background and injections is not needed. For the case of large required significance, the sensitivity of the \pycbc{} search is overestimated (up to factor of two) when higher-order modes are omitted in the injected signals. 

The upper limits on coalescence rates of IMBHB presented in \cite{Abbott:2017iws} are quoted at a significance threshold corresponding to an inverse false alarm rate $< 1yr$. We have shown that, under this condition, omission of higher modes in the injection sets leads to an underestimation of the sensitivity of both searches for the case of highly asymmetric binaries. Hence, the corresponding upper limits are fairly conservative. However, the most stringent upper limits, which were placed for nearly equal-mass binaries, are robust with respect to the omission of higher modes.

We also compared the sensitivity of the two considered searches. We find that systems heavier than $200M_{\odot}$, can be confidently detected by $\verb+cWB+$ within a larger volume than \pycbc{}. The increase is due in part to the impact of higher-modes and in part to the different response of each search to the non-stationary instrumental noise. In fact, when higher-modes are omitted (included), the increase is a factor of 5 (10). When relaxing the threshold of confidence, however, the matched-filter search outperforms the un-modeled search across the entire parameter space here considered.

We stress here that we have run both searches in their Advanced LIGO O1 Science run configuration. The second observing run, recently concluded, employed different configurations for both search pipelines considered here \cite{Nitz:2017svb, DalCanton:2017ala,Nitz:2017lco}. Some of these improvements targeted specifically the high background associated with short, high-mass binary merger templates. The results we presented here might therefore not apply to the updated configurations and it would be interesting to revisit them in a future study.

In addition, there is an ongoing effort to include higher-order modes in the templates of the \pycbc{} search \cite{Harry:2017weg}. This should improve the sensitivity of \pycbc{} toward highly asymmetric binaries. Consequently, the results presented in this work will need to be re-evaluated once such a search is ready. Finally, an interesting extension of this study will be checking the ability of both described searches to recover signals from spinning binary black holes including higher-order modes, especially in the case of precessing systems.

\section{Acknowledgements}

We thank Alex Nitz and Henning Fehrmann for their help with the \pycbc{} pipeline. JCB and KJ gratefully acknowledge support from the NSF grants 1505824, 1505524, 1333360. TDC was supported by an appointment to the NASA Postdoctoral Program at the Goddard Space Flight Center, administered by Universities Space Research Association under contract with NASA. Computations used in this work were performed on the Atlas high-throughput computing clusters operated by the Max Planck Institute for Gravitational Physics.

\section{Appendix: Testing the EOBNRv2HM waveform model}

\begin{figure*}[htb!]
\centering
\includegraphics[width=1.\columnwidth]{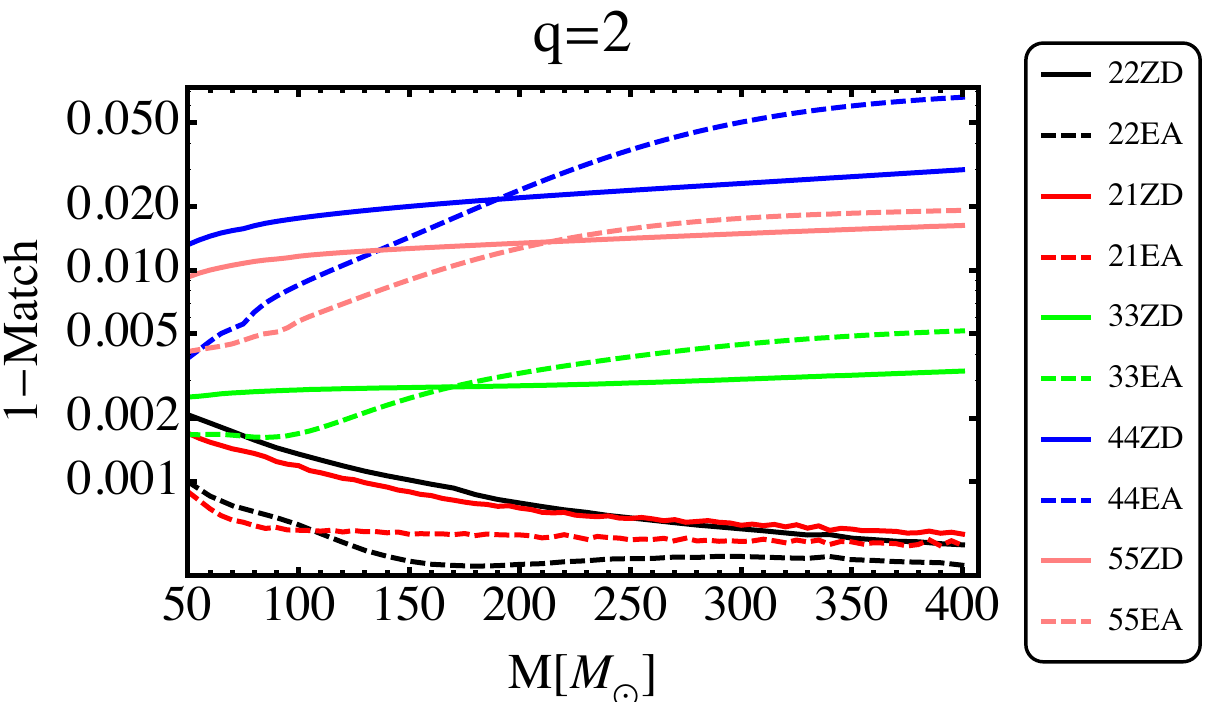}
\includegraphics[width=1.\columnwidth]{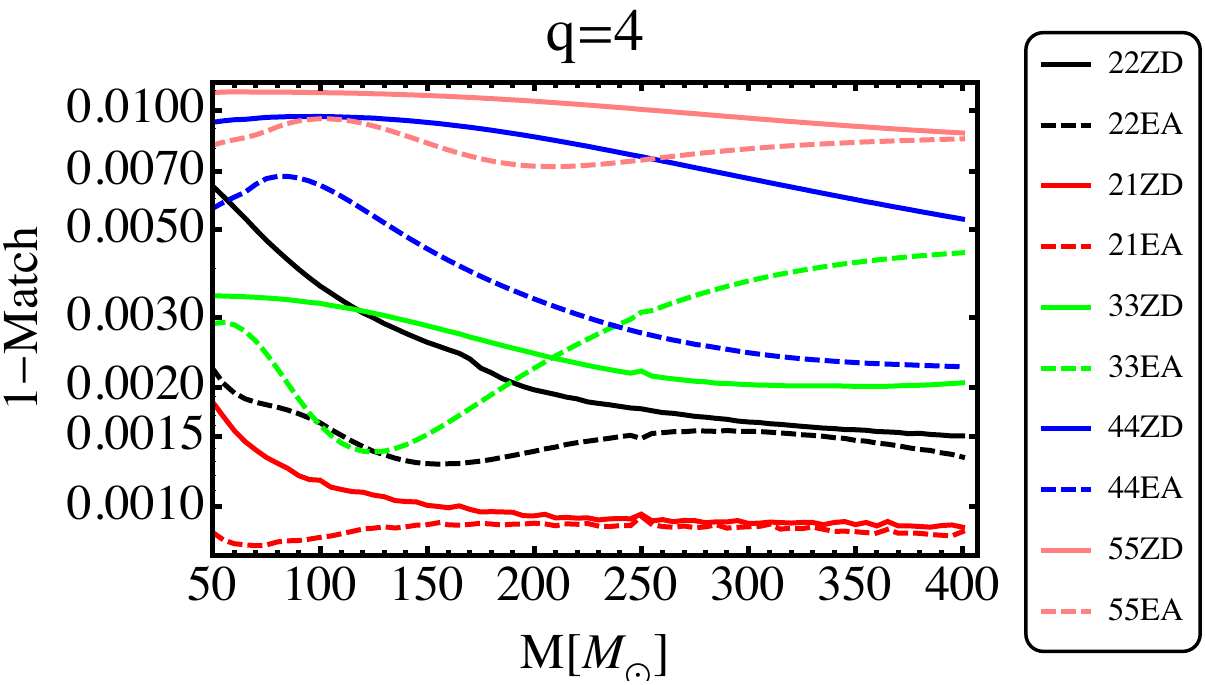}
\includegraphics[width=1.\columnwidth]{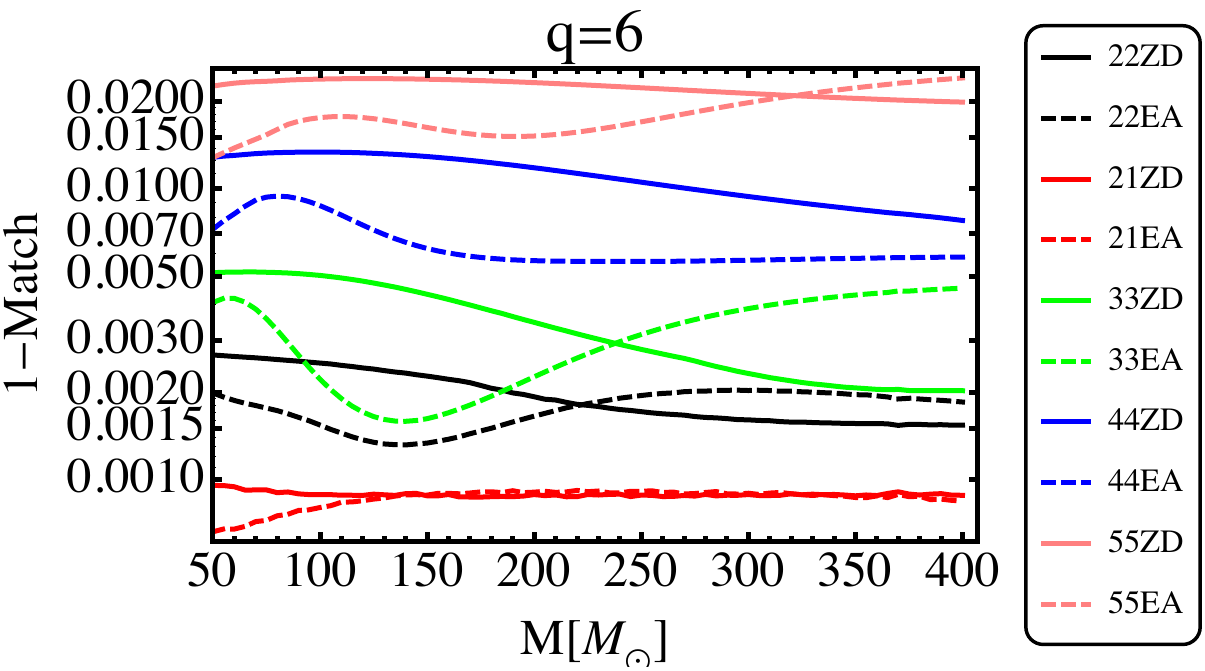}
\includegraphics[width=1.\columnwidth]{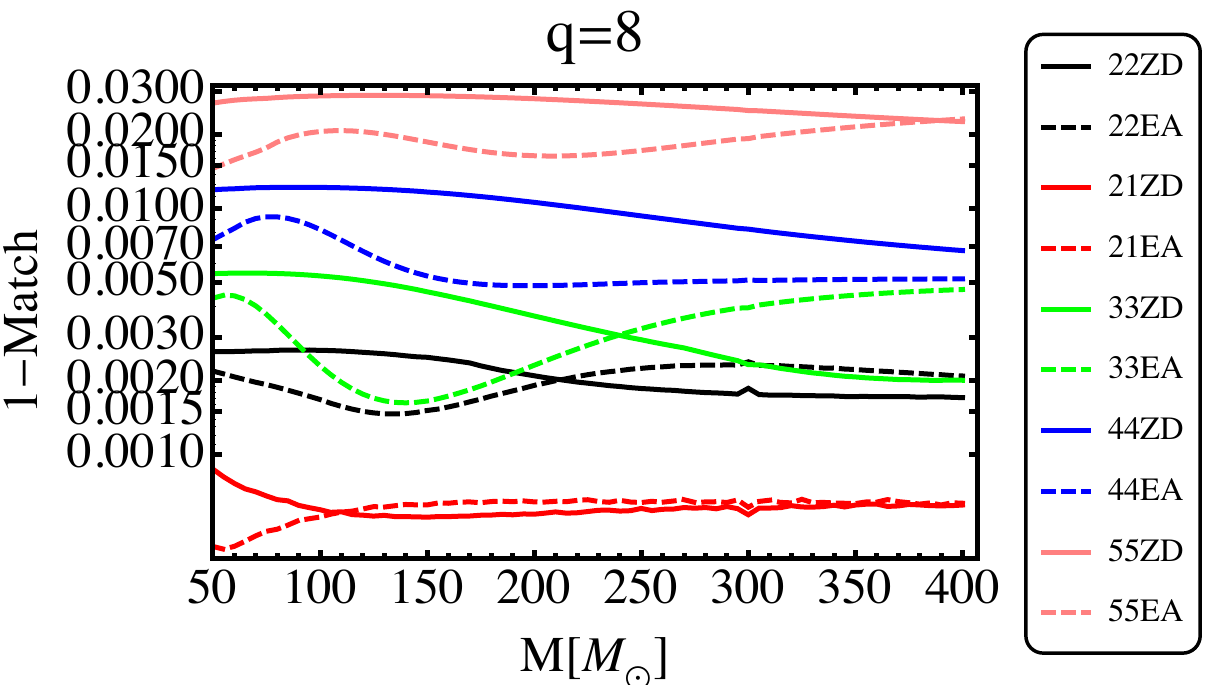}
\caption[Effect of $q$ on higher modes]{Mismatch, or unfaithfulness, of the modes computed by  EOBNRv2HM towards the corresponding SXS Numerical Relativity ones. The noise curve used is early Advanced LIGO (EA) with $f_{low}=20$Hz for the dashed lines and Zero-detuned High-energy-power (ZD) with $f_{low}=10$Hz for the solid ones.}
\label{ex:fig:EOBhigher-modes}
\end{figure*}

\begin{figure*}[htb!]
\centering
\includegraphics[width=1.\columnwidth]{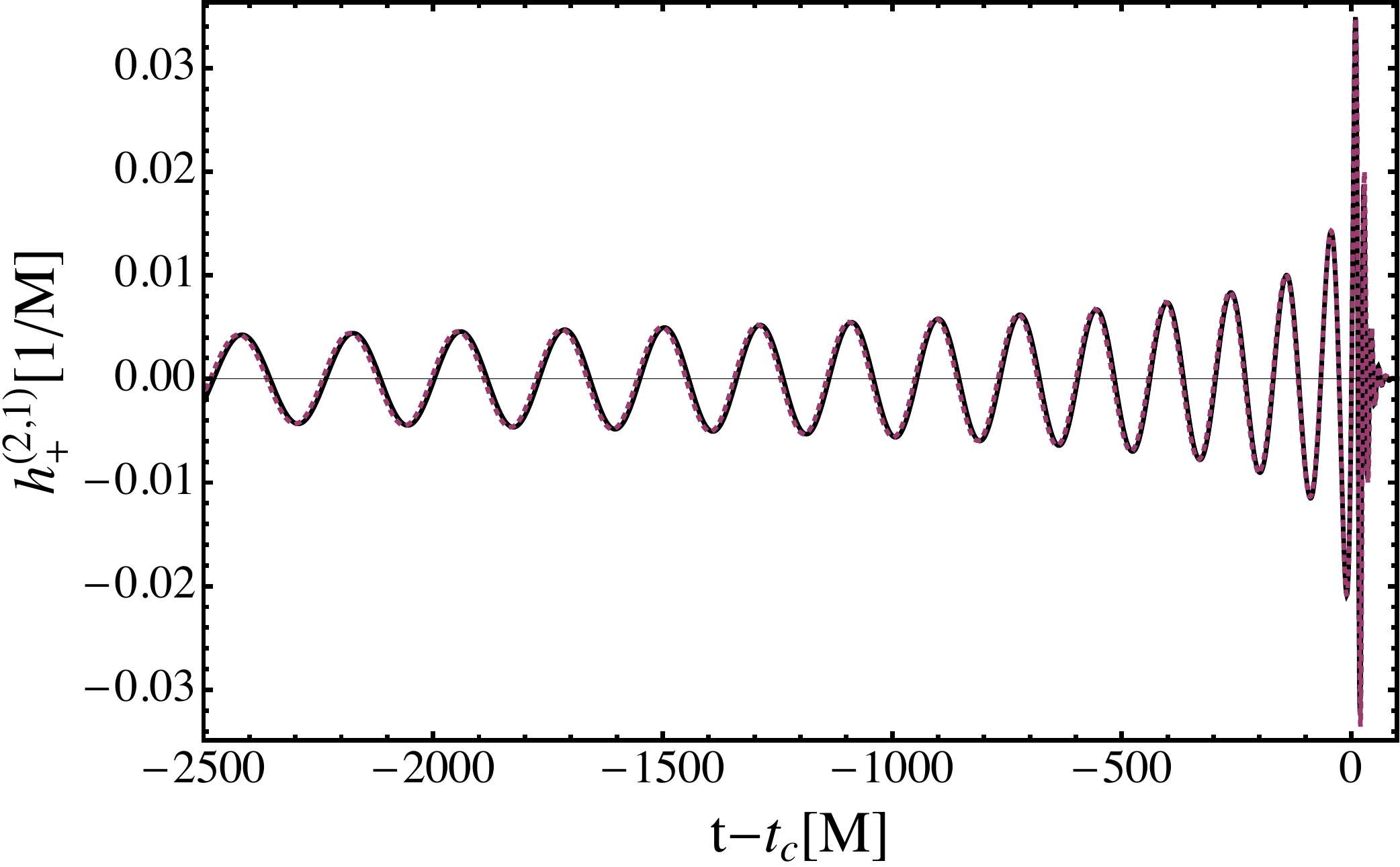}
\includegraphics[width=1\columnwidth]{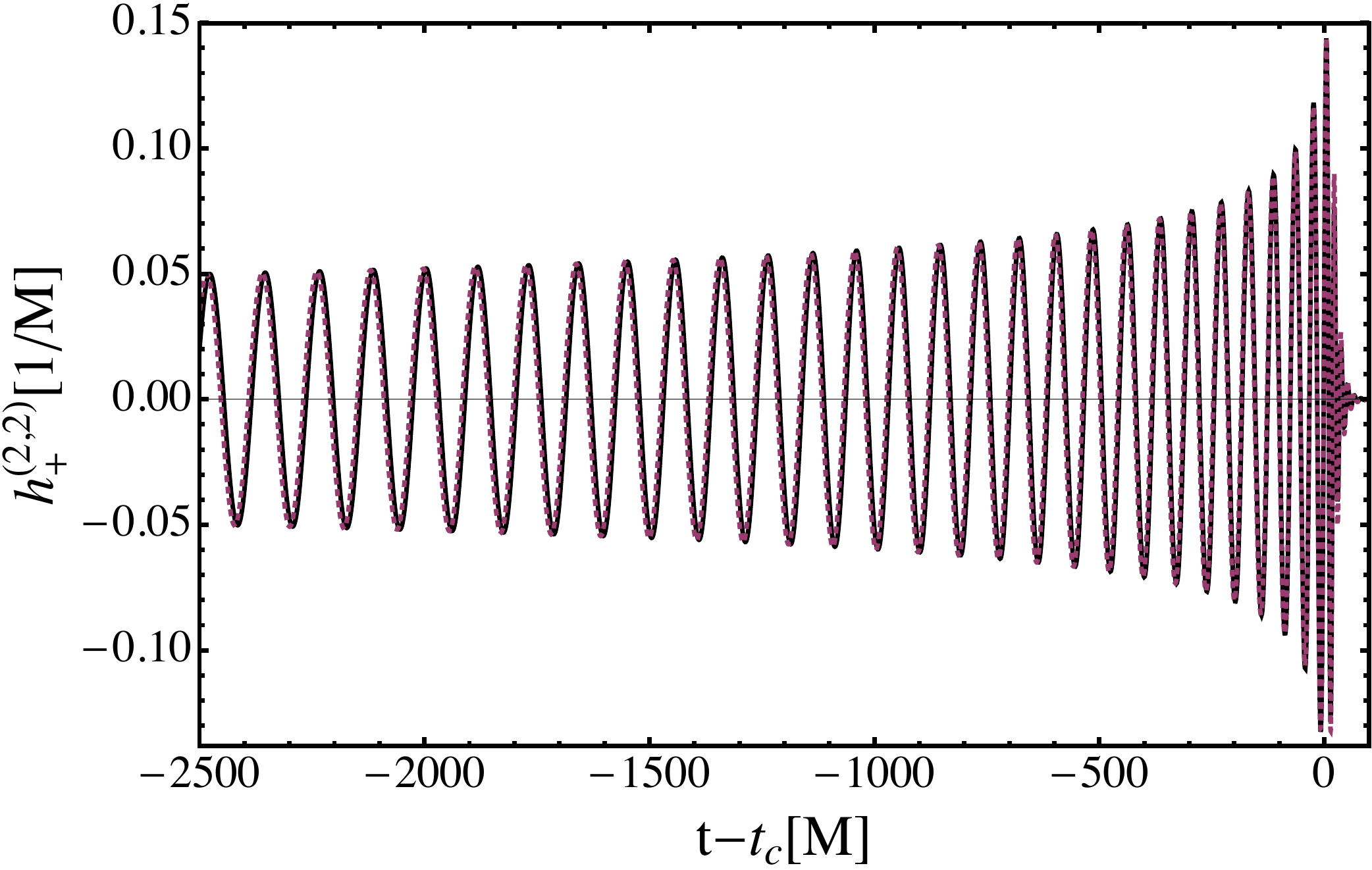}
\includegraphics[width=1\columnwidth]{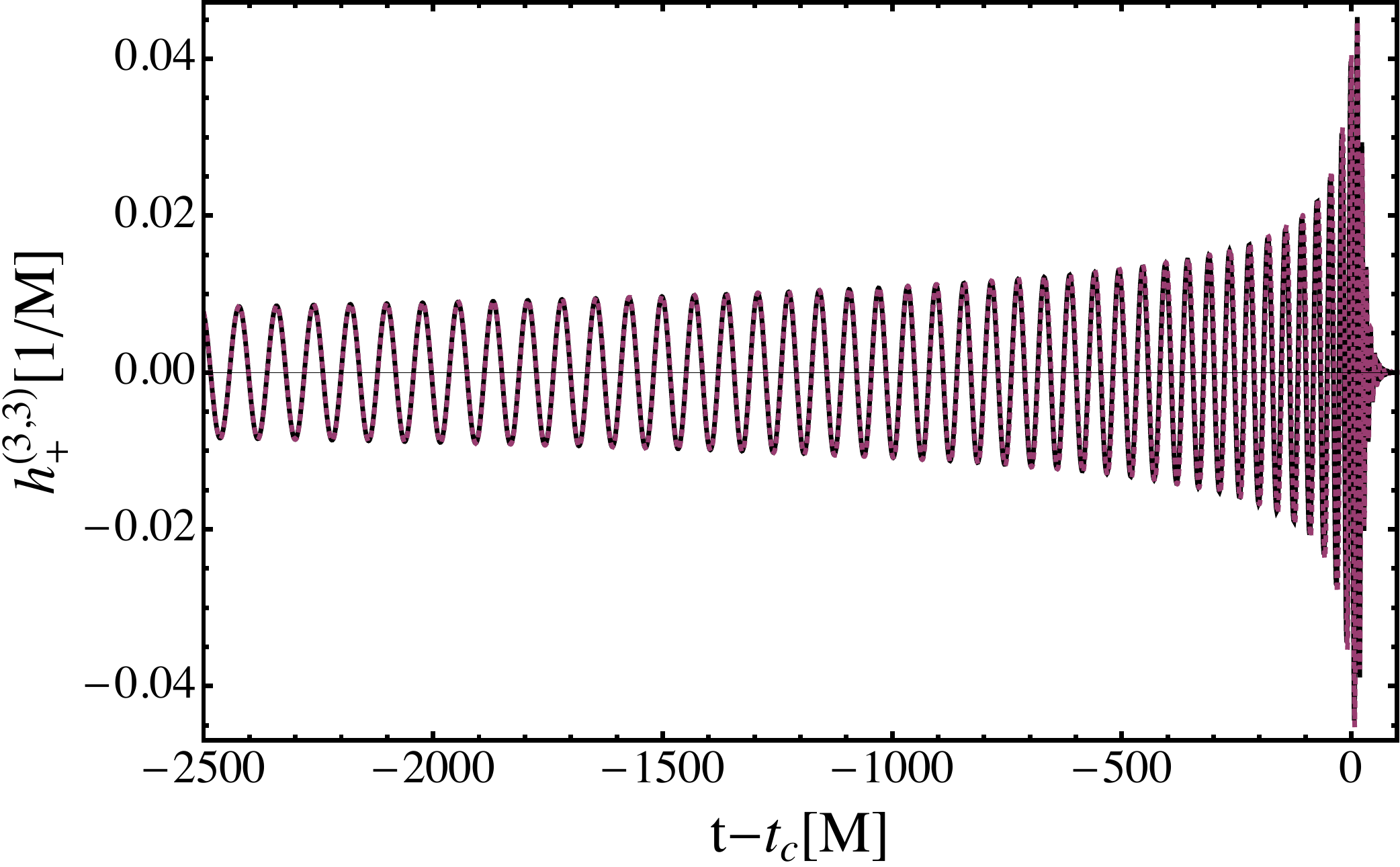}
\includegraphics[width=1\columnwidth]{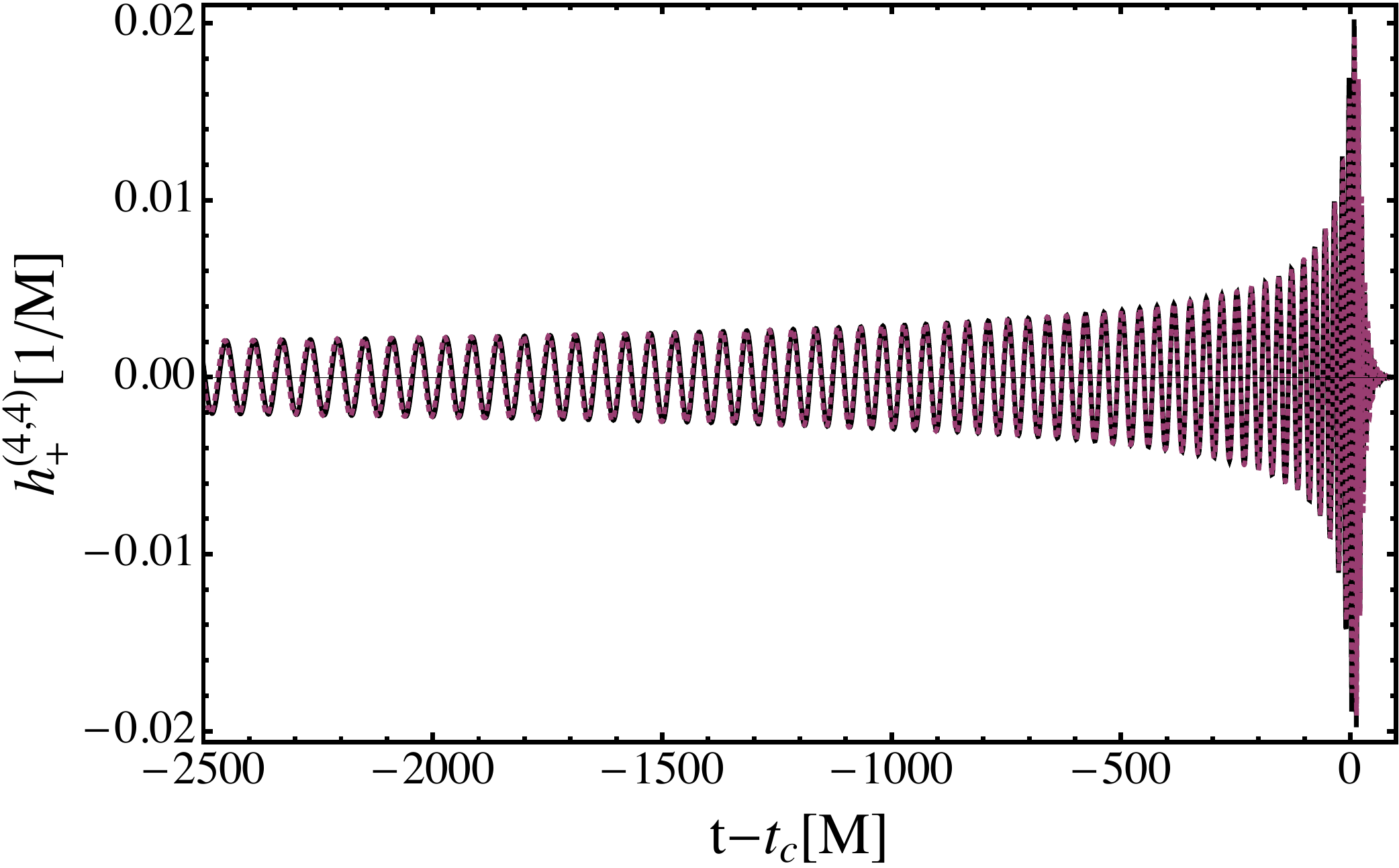}
\caption[Effect of $q$ on higher modes]{Last cycles of EOBNRv2HM (Black) and SXS modes (Red) for the case of a $q=8$ binary. The SXS simulation is SXS:BBH:0063, the axes are in geometrical units and $t_c$ denotes the coalescence time, which we make coincide with the peak of the $(2,2)$ mode.}
\label{ex:fig:modescomparison}
\end{figure*}

We dedicate this appendix to assess the accuracy of the $\verb+EOBNRv2HM+$ model. In order to accomplish this, we compute the overlap of the $\verb+EOBNRv2HM+$ modes, $h_{\ell,m}^{EOB}$, to the corresponding modes computed by means of numerical relativity (NR), $h_{\ell,m}^{NR}$. In particular, we choose a set of NR non-spinning waveforms computed by the SpEC code \cite{SXS,Hemberger:2012jz,Buchman:2012dw,Szilagyi:2009qz}, made public by the SXS collaboration \cite{SXS}. Fig. \ref{ex:fig:EOBhigher-modes} shows the overlap, optimised over time and phase offsets of the EOB modes towards the SXS ones. This is:

\begin{equation}
(h^{EOB}_{\ell,m} | h^{NR}_{\ell,m})=\max_{\Delta t, \Delta \phi}\frac{\langle h^{EOB}_{\ell,m} | h^{NR}_{\ell,m} \rangle}{\sqrt{\langle h^{EOB}_{\ell,m} | h^{EOB}_{\ell,m} \rangle \langle h^{NR}_{\ell,m} | h^{NR}_{\ell,m} \rangle}}
\end{equation}

Most mismatches between the $(2,2)$ modes are lower than 0.005 except for the case of $q=4$, which gets to 0.007 in the low mass end. The worst overlaps are obtained for either low mass ratios (see the $(4,4)$ mode for $q=2$) (for which higher-modes are negligible) or for high $m$ modes, particularly the $(5,5)$, which is the least dominant of them. For the rest of dominant modes, matches are always larger than $0.985$ except for the $q=2$ case. In particular, matches between the most dominant higher-modes, namely the $(3,3)$, are always larger than 0.995.\\

Since the $\verb+EOBNRv2HM+$ model was calibrated to SXS waveforms, these large matches should not be surprising. In order to have an independent cross check, we have also calculated the matches of the $\verb+EOBNRv2HM+$ modes to numerical relativity modes produced by the $\verb+MAYA+$ code, available in the Georgia Tech catalogue of numerical waveforms \cite{Jani:2016wkt,Bustillo:2016gid}. Although not all the waveforms used in Fig.\ref{ex:fig:EOBhigher-modes} are available in this catalogue, we could compare the $q=2$ and $q=6$ cases. Similar results to the ones described for SXS were obtained with the exception of the $(4,4)$ mode of the  $q=6$ case. All $\ell < 3$ modes yielded matches larger than 0.985, while the others yielded results above 0.95.

We stress that these matches have been independently optimised over time and phase shifts for each separate mode. Doing this, however, might hide potential phase offsets $\epsilon_{\ell,m}$ between the individual EOB and SXS modes, as described in \cite{Bustillo:2015ova}. In other words, when computing 5 independent matches for the 5 $(\ell,m)$ modes, we are optimising over 10 independent degrees of freedom: 5 time offsets $\Delta t_{\ell,m}$ and 5 phase shifts $\Delta \phi_{\ell,m}$. However, the time offset should be equal for all the modes and the phase offset should satisfy $\Delta \phi_{\ell,m} = \frac{m}{2} \Delta \phi_{2,2}=m \Delta\phi_{orb}$, where $\Delta\phi_{orb}$ denotes the differences in the orbital phase in the EOB and NR calculations at a given frequency. Denoting by $\Delta \phi_{\ell,m}$ the phase shift needed for optimising the match between the $(\ell,m)$ modes, $\epsilon_{\ell,m}$ are defined as \cite{Bustillo:2015ova}:
\begin{equation}
\epsilon_{\ell,m}=\Delta \phi_{\ell,m} - \frac{m}{2} \Delta \phi_{2,2}.
\end{equation}
Non-zero values indicate that NR and $\verb+EOBNRv2HM+$ have different phase offsets wrt., the corresponding $(2,2)$ modes, which will lead to different resulting full waveforms, even if the optimised match of the individual modes is $1$. This is, after accounting for the orbital phase shift $\Delta \phi_{orb}$ inferred from $\Delta \phi_{2,2}$, one would obtain two different complex strains $\cal H$ for a given orientation of the binary:

\begin{equation}
\begin{aligned}
& {\cal H}^{NR}=\sum_{\ell\geq 2}\sum_{m=-\ell}^{m=\ell}Y^{-2}_{\ell,m}(\theta,\varphi)h_{\ell,m}(\Xi;t)\\ &
{\cal H}^{EOB}=\sum_{\ell\geq 2}\sum_{m=-\ell}^{m=\ell}Y^{-2}_{\ell,m}(\theta,\varphi)h_{\ell,m}(\Xi;t)e^{-i\epsilon_{\ell,m}}
\end{aligned}
\end{equation}
where $\epsilon_{2,2}=0$ by definition and 
\begin{equation}
(h_{\ell,m} | h_{\ell,m} e^{-i\epsilon_{\ell,m}})=1.
\end{equation}

The $\epsilon_{\ell,m}$ of the NR modes relative to the $\verb+EOBNRv2HM+$ show values of some tenths of a degree in the worst case. In \cite{Bustillo:2015ova} shows that values ten times larger than these produce mismatches between the full waveforms below $\sim 0.01$. We thus conclude that the $\verb+EOBNRv2HM+$ model is accurate enough for our purposes.\\ 

Finally, in order to allow for a visual comparison, Fig.\ref{ex:fig:modescomparison} shows the $+$ polarization of the $(\ell,m)$ modes predicted by the Numerical Relativity SpEC code and $\verb+EOBNRv2HM+$ higher-modes for a $q=8$ binary, in geometric NR units. The time alignment of the modes has been performed such that the peak of the $(2,2)$ mode occurs at $t_c=0$, which stands for coalescence time.

\bibliography{HMbib.bib}

\end{document}